\documentclass[12pt]{article}

\usepackage[left=2.50cm,right=2.50cm,top=2.50cm,bottom=2.50cm]{geometry}
\usepackage[utf8]{inputenc}
\usepackage[english]{babel}
\usepackage{amsmath,amssymb,mathtools}
\usepackage{graphicx}
\graphicspath{{./figures/}{./}}
\usepackage{booktabs,array}
\usepackage{cite}
\usepackage{float}
\usepackage[hang,flushmargin]{footmisc}
\usepackage{hyperref}
\usepackage{caption}
\hypersetup{hidelinks}
\numberwithin{equation}{section}

\newcommand{\be}{\begin{equation}}
\newcommand{\ee}{\end{equation}}
\newcommand{\dd}{\mathrm d}
\newcommand{\IR}{\mathbb R}
\newcommand{\M}{\mathcal M}
\newcommand{\Acal}{\mathcal A}
\newcommand{\Eint}{E_{\rm int}}
\newcommand{\Tr}{\operatorname{Tr}}

\newlength{\bibitemsep}
\let\oldthebibliography\thebibliography
\renewcommand\thebibliography[1]{%
  \oldthebibliography{#1}%
  \raggedright
  \setlength{\parskip}{\bibitemsep}%
  \setlength{\itemsep}{0pt}%
}

\begin{document}

\noindent
{\LARGE\bf From photons to fractons: Casimir effect in higher-rank electrodynamics and its fractonic branch}

\vspace{0.2cm}
\hrule
\vspace{0.45cm}
\noindent
{\bf Nicola Maggiore$^{1,2}$}
\par\vspace{0.25cm}

\noindent$^1$ Dipartimento di Fisica, Universit\`a di Genova, Via Dodecaneso 33, I-16146 Genova, Italy.

\noindent$^2$ Istituto Nazionale di Fisica Nucleare -- Sezione di Genova, Via Dodecaneso 33, I-16146 Genova, Italy.

\vspace{0.35cm}
\noindent{\tt Abstract}\\
The Casimir effect probes the modes that can propagate between boundaries. This makes it useful for fracton gauge theories, where restricted mobility follows from the simultaneous conservation of charge and dipole moment and changes the physical sector of the theory. We compare ordinary Maxwell electrodynamics with a Lorentz-covariant higher-rank theory of a symmetric tensor gauge field and with the scalar-charge fractonic branch selected from it by a preferred timelike background. The comparison shows why the distinction matters. In the full covariant theory, the Euclidean slab problem contains three independently propagating Maxwell-like modes and two coupled vector sectors that do not reduce to independent ordinary modes. Their boundary determinants are well defined, but a Lorentzian analysis shows that the vector sector has an unbounded Hamiltonian, so these determinants do not describe the vacuum energy of a stable fractonic phase. On the fractonic branch, the Gauss constraint removes that vector sector and leaves three stable massless modes: two transverse-traceless tensor modes and one scalar trace mode. For ideal reflectors, the physical Casimir energy is therefore 3/2 of the corresponding Maxwell result, for both equal and complementary boundary conditions. The calculation isolates the effect of the fractonic restriction on the radiative spectrum and clarifies which features of the full higher-rank boundary problem survive in the physical fracton sector.\par\vspace{0.3cm}
\noindent{\tt Keywords:}\\
Casimir effect; fractons; symmetric tensor gauge theory; Jordan blocks; quantum field theory with boundaries.

\vspace{0.35cm}
\hrule
\vspace{0.2cm}
{\raggedright\noindent{\tt E-mail: nicola.maggiore@ge.infn.it.}\par}

\newpage

\section{Introduction}
\label{section_introduction}

The Casimir effect is a change of the quantum vacuum caused by boundaries \cite{Casimir:1948dh,Lifshitz:1956zz,Milton:2001yy,Bordag:2001qi,Bordag:2009zz}. For two parallel plates, the physically relevant quantity is the part of the vacuum energy that depends on their separation. Contributions that remain when the plates are infinitely far apart---the empty-space term and the self-energy of each isolated plate---are subtracted. In scattering language, the remaining interaction is built from propagation across the gap and reflection at the two surfaces \cite{Lambrecht:2006few,Emig:2007cf,Rahi:2009hm,Reynaud:2010vr}. This makes the Casimir effect a spectral probe: it depends on which modes can travel between the plates and on how the boundaries act on those modes.

That observation is especially relevant for fractons. Fracton phases are characterized by restricted mobility, which in the simplest scalar-charge theory follows from conservation of both total charge and total dipole moment \cite{Nandkishore:2018sel,Pretko:2020cko,Pretko:2016lgv,Pretko:2016kxt,Seiberg:2020wsg}. If an isolated charge were displaced, the dipole moment would change; the simultaneous conservation laws therefore forbid such a motion. Neutral composites, such as dipoles, can remain mobile. In the gauge-theory description this structure is encoded by a symmetric spatial tensor $A_{ij}$, a scalar gauge parameter $\Lambda$, and the associated higher-rank gauge symmetry \cite{Pretko:2016lgv,Pretko:2016kxt}. The gauge transformation is
\be
\delta A_{ij}=\partial_i\partial_j\Lambda,
\label{intro_scalar_charge_gauge}
\ee
and a Gauss law with two spatial derivatives,
\be
\partial_i\partial_j E^{ij}=\rho.
\label{intro_scalar_charge_gauss}
\ee
Here $E^{ij}$ is the gauge-invariant electric tensor and $\rho$ is the scalar charge density. Integrating Eq.~\eqref{intro_scalar_charge_gauss} over space gives charge conservation; multiplying it by a spatial coordinate before integrating gives dipole-moment conservation, up to the usual boundary terms. The same double-derivative structure therefore explains both the use of a symmetric tensor gauge field and the defining mobility restriction of the scalar-charge fracton phase.

Casimir-like interactions have already been discussed in a different fractonic setting. In Ref.~\cite{Feng:2022fractoncasimir}, Hilbert-space fragmentation in a one-dimensional charge- and dipole-conserving system produces an effective attraction between fractons, or between a fracton and a boundary, in analogy with the Casimir effect. That mechanism is not a vacuum interaction generated by fluctuations of a higher-rank gauge field. The present problem instead concerns the vacuum fluctuations of a propagating higher-rank gauge field in the presence of two boundaries.

A Lorentz-covariant extension promotes the spatial tensor to a symmetric spacetime field $A_{\mu\nu}=A_{\nu\mu}$ and the gauge transformation to
\be
\delta A_{\mu\nu}=\partial_\mu\partial_\nu\Lambda.
\label{intro_tensor_gauge}
\ee
The corresponding Maxwell-like action was introduced in Ref.~\cite{Bertolini:2022ijb}, within the wider family studied in Ref.~\cite{Blasi:2022mbl}; related gauge-fixing, BV--BRST and higher-dimensional formulations appear in Refs.~\cite{Bertolini:2023gauging,Fecit:Rovere:2026worldline,Maggiore:2026sixdim}. Other covariant higher-rank constructions displaying fractonic behavior include three-dimensional BF-like and quasi-topological models \cite{Bertolini:2025bf,Bertolini:2025quasi}, while a spacetime-dipole formulation with longitudinal diffeomorphisms was developed in Ref.~\cite{Afxonidis:2024spacetime}. This full Lorentz-covariant theory is broader than the scalar-charge fracton sector. Throughout the paper, the expressions \emph{full covariant theory} and \emph{unrestricted higher-rank theory} refer to this same model; we alternate them only to avoid repetitive terminology. The scalar-charge sector, by contrast, requires a preferred notion of time, as expected from its nonrelativistic physical interpretation. A convenient tensorial formulation introduces a constant timelike background vector that identifies the rest frame of the reduced theory. In that frame the restriction takes the component form $A_{\mu0}=\partial_\mu\phi$, where $\phi$ is a scalar. This geometric distinction between the full covariant theory and the preferred-frame scalar-charge reduction is central to the present analysis. While this work was being completed, a preprint appeared on the zero- and finite-temperature Casimir effect of the autonomous trace-full scalar-charge gauge theory \cite{Muniz:2026fractoncasimir}, which varies that theory directly and finds five polarizations, three of speed $v$ and two of speed $v/\sqrt2$, with a zero-temperature ratio $(3+\sqrt2)/2$ to ideal electromagnetism. The problem solved here is different, and the relation between the two is exact. The reduced action is the same functional, but it is not varied autonomously: the branch inherits the equations of the covariant theory, among them the vector constraint $\partial_iE^{ij}=0$, which implies the scalar law of Eq.~\eqref{intro_scalar_charge_gauss} but is strictly stronger. Its transverse projection removes the two transverse-vector amplitudes, which are precisely the two slow modes of the autonomous theory, leaving three modes with the common propagation speed of the covariant equations. The two coefficients therefore differ by exactly the slow-branch contribution, $(3+\sqrt2)/2-3/2=\sqrt2/2$: the results do not conflict, and their difference measures what the covariant embedding removes.

The physical question of this paper is then simple to state. If one computes a Casimir interaction using the full higher-rank theory, does one obtain the Casimir effect of the fractonic phase, or does the restriction to the scalar-charge branch change the propagating spectrum and therefore the force? To answer this question without mixing distinct effects, we compare three systems in the same slab geometry: ordinary Maxwell electrodynamics, the full covariant higher-rank theory, and its fractonic branch. Maxwell provides the standard benchmark. The full tensor theory shows what the covariant completion adds. The branch calculation then identifies which part of that structure survives in the covariantly embedded fractonic sector.

Our method follows the standard boundary-field-theory logic. We first formulate the vacuum problem in Euclidean frequency, where propagation across the gap is exponentially damped and the Casimir interaction can be written as a finite determinant. For Maxwell theory the two physical photon polarizations provide two independent channels. For the higher-rank theory we derive the normal modes, fix the scalar gauge symmetry in a Becchi--Rouet--Stora (BRS)-invariant way, and obtain the boundary conditions from local surface actions following Symanzik's approach \cite{Symanzik:1981wd,Vassilevich:2003xt,Amoretti:2014boundary}. Closely related boundary-field-theory constructions have been used for Maxwell and Maxwell--Chern--Simons systems \cite{Maggiore:2018mcs,Bertolini:2020holographic,Maggiore:2026agw}. The subtraction at infinite plate separation removes bulk and one-plate factors, leaving only the interaction that connects the two surfaces. Gauge-parameter independence is checked both algebraically and through the explicit cancellation of the longitudinal and Faddeev--Popov determinants.

The full covariant theory has a richer Euclidean spectrum than Maxwell theory. Besides three independently propagating Maxwell-like amplitudes, it contains two coupled vector pairs. Each pair has repeated characteristic roots but only one ordinary eigenvector at each root; generalized solutions are therefore required. We refer to such a coupled pair as a \emph{Jordan doublet}. The associated finite determinants depend on how the two components couple to the plates, and we compute them for fixed, free, mixed and orientation-sensitive boundary conditions. These results are mathematically well-defined Euclidean interaction functionals. Their physical interpretation, however, requires a separate Lorentzian check.

That check is decisive. The same vector pairs that produce the Jordan structure in the slab form a dynamical sector whose Hamiltonian is not bounded below. The full covariant theory therefore does not provide a stable Lorentzian vacuum whose zero-point energy could be identified with the Casimir energy of a physical fracton phase. We keep its Euclidean determinants because they reveal which spectral and boundary features belong to the unrestricted covariant completion, but we distinguish them throughout from the physical Casimir energy.

The fractonic branch behaves differently. The preferred-frame restriction must be imposed together with the Gauss equations inherited from the full theory. In the source-free quadratic theory these constraints eliminate the transverse-vector sector and fix the longitudinal scalar, leaving three stable massless radiative modes: two transverse-traceless tensor polarizations and one scalar trace mode. This constraint structure implies the source-free form of the scalar-charge Gauss law in Eq.~\eqref{intro_scalar_charge_gauss}, which is responsible for charge and dipole conservation when matter is present, but is not equivalent to it: the inherited condition is a vector equation, while the scalar-charge law is its double divergence alone. The Casimir problem therefore connects directly to the defining fracton physics: the restriction that defines the embedded fractonic branch also changes the set of radiative modes available between the plates.

For ideal reflectors the consequence is particularly transparent. Maxwell theory has two stable physical channels, whereas the fractonic branch has three. After the reduced Hamiltonian and boundary problem are derived, the branch Casimir energy is found to be exactly three halves of the corresponding Maxwell value, for both equal and complementary ideal boundary conditions. The result is not assumed from component counting: it follows only after the Gauss constraints have removed the unstable vector sector and the three remaining modes have been shown to propagate independently with positive energy. By contrast, the larger numerical coefficients obtained from the full covariant theory arise from the coupled Jordan sector and are retained only as Euclidean spectral diagnostics.

The new physical information is therefore not exhausted by the numerical factor relating the two Casimir energies. The same restriction that realizes scalar-charge fracton kinematics also removes the Lorentzian-unstable vector sector that is responsible for the Jordan enhancement of the unrestricted Euclidean boundary problem. The Casimir comparison thus separates, in a directly observable vacuum quantity, the radiative content of the stable fractonic branch from spectral structures that belong only to the larger covariant theory.

The boundary analysis also exposes a second distinction. Boundary conditions derived for the full tensor field cannot in general be imposed unchanged after the fractonic restriction, because components that were independent before the reduction become related by the Gauss constraints. For the physical branch we therefore define ideal reflectors directly on the reduced three-mode space. Constructing more general local surface actions compatible with the preferred-frame reduced phase space is a separate problem and is not replaced here by arbitrary phenomenological reflection coefficients.

The paper is organized as follows. Section~\ref{section_maxwell} establishes the common slab geometry and the Maxwell benchmark. Section~\ref{section_tensor} introduces the full covariant higher-rank theory and its BRS gauge fixing. Section~\ref{section_modes} derives its normal-mode decomposition and analyzes the Lorentzian stability of the coupled vector sector. Section~\ref{section_boundary_doublets} develops the local surface response and the associated Euclidean interaction determinants. Section~\ref{section_fractonic_branch} constructs the scalar-charge branch relative to a timelike background, derives its physical spectrum and formulates ideal branch-compatible reflectors. Section~\ref{section_results} compares the three systems. Section~\ref{section_odd} studies orientation-dependent surface response in the full covariant theory, and Section~\ref{section_conclusions} summarizes the physical implications.

\section{Common slab geometry and the Maxwell benchmark}
\label{section_maxwell}

\subsection{Geometry and notation}

We use the mostly-plus Lorentzian metric
\be
\eta_{\mu\nu}=\operatorname{diag}(-,+,+,+).
\label{metric}
\ee
The two parallel plates are located at $x^3=0$ and $x^3=h$. The region between them is the slab
\be
\M=\IR^{1,2}\times[0,h],
\qquad
\partial\M=\Sigma_0\cup\Sigma_h,
\label{slab}
\ee
where $h>0$ is the plate separation and $\Sigma_\alpha$, with $\alpha\in\{0,h\}$, denotes the plate at $x^3=\alpha$. We denote by $\Acal$ the plate area and quote all vacuum energies per unit area; for infinite parallel plates, $\Acal$ is understood as the usual transverse normalization area. The outward unit normals are
\be
n_\mu^{(0)}=-\delta_\mu^{\ 3},
\qquad
n_\mu^{(h)}=+\delta_\mu^{\ 3}.
\label{normals}
\ee
They define the projector onto directions tangent to each plate and the orientation tensor induced on it,
\be
{\cal P}^{(\alpha)}_{\mu\nu}
=\eta_{\mu\nu}-n^{(\alpha)}_\mu n^{(\alpha)}_\nu,
\qquad
\varepsilon^{(\alpha)}_{\mu\nu\rho}
=n_{(\alpha)}^\sigma\varepsilon_{\sigma\mu\nu\rho}.
\label{boundary_geometry}
\ee

The vacuum energy is most conveniently evaluated after rotating the frequency integration to imaginary frequency, $t=-i\tau$. This standard Wick rotation turns the free wave operator into an elliptic Euclidean operator and replaces oscillatory propagation across the gap by exponentially decaying factors; it is the usual starting point of determinant and scattering derivations of the Casimir effect \cite{Milton:2001yy,Bordag:2001qi,Bordag:2009zz,Rahi:2009hm}. We Fourier transform the three coordinates parallel to the plates, $x^a=(\tau,x^1,x^2)$, and write
\be
p_a=(\zeta,k_1,k_2),
\qquad
q\equiv |p|=\sqrt{\zeta^2+k_1^2+k_2^2},
\qquad
X\equiv qh.
\label{tangential_momentum}
\ee
Here $\zeta$ is the Euclidean frequency, $(k_1,k_2)$ is the spatial momentum parallel to the plates, $q$ is the modulus of the three-dimensional Euclidean tangential momentum, and $X$ is the corresponding dimensionless separation variable.

\subsection{Physical Maxwell modes}

The Euclidean action of ordinary Maxwell electrodynamics is
\be
S_{\rm EM}=\frac{g_{\rm EM}}4\int_{\M}\dd^4x\,F_{\mu\nu}F_{\mu\nu},
\qquad
F_{\mu\nu}=\partial_\mu A_\nu-\partial_\nu A_\mu,
\label{maxwell_action}
\ee
where $A_\mu$ is the electromagnetic gauge potential and $g_{\rm EM}>0$ is an overall normalization. Because the same constant multiplies the quadratic operator at finite $h$ and at $h\to\infty$, it cancels from the normalized determinant used for the interaction energy.

At fixed nonzero $p_a$, choose two orthonormal polarization vectors $e_a^{(s)}$, $s\in\{{\rm TE},{\rm TM}\}$, where TE and TM denote transverse-electric and transverse-magnetic polarization, respectively. They satisfy
\be
p^a e_a^{(s)}=0,
\qquad
e^{(s)a}e_a^{(s')}=\delta_{ss'}.
\label{maxwell_polarization_basis}
\ee
The two physical tangential amplitudes can then be written as
\be
A_a^{\perp}(p,x^3)
=e_a^{({\rm TE})}\phi_{\rm TE}(x^3)
+e_a^{({\rm TM})}\phi_{\rm TM}(x^3).
\label{maxwell_TE_TM_decomposition}
\ee
After continuation back to real frequency, these become the physical TE and TM photon polarizations. The remaining components are fixed by the Maxwell constraint and gauge freedom. Substituting Eq.~\eqref{maxwell_TE_TM_decomposition} into the free Maxwell equations shows that the two physical amplitudes obey the same normal equation,
\be
\left(-\partial_3^2+q^2\right)\phi_s=0,
\qquad s\in\{{\rm TE},{\rm TM}\}.
\label{intro_helmholtz}
\ee
Thus a free Maxwell polarization propagates across the Euclidean slab through the two profiles $e^{\pm qx^3}$ \cite{Milton:2001yy,Bordag:2001qi,Bordag:2009zz}. We shall call an independent higher-rank mode \emph{Maxwell-like} when its normal amplitude obeys the same second-order equation.

For one such amplitude, the quadratic bulk action at fixed $p_a$ is proportional to
\be
S^{\rm bulk}_s
=\frac{g_{\rm EM}}2\int_0^h\dd x^3\,
\left(|\phi_s'|^2+q^2|\phi_s|^2\right),
\label{maxwell_mode_action}
\ee
where a prime denotes $\partial_3$. A local quadratic surface term on plate $\Sigma_\alpha$ can be written
\be
S^{(s)}_{\partial,\alpha}
=\frac{g_{\rm EM}}2\,K_{\alpha,s}(q)\,|\phi_s|^2_{\Sigma_\alpha}.
\label{maxwell_surface_action}
\ee
The real function $K_{\alpha,s}(q)$ characterizes the response of that plate to polarization $s$; allowing it to depend on $q$ includes a dispersive but translation-invariant response along the surface. Varying $S^{\rm bulk}_s+S^{(s)}_{\partial,\alpha}$ gives the Robin boundary condition
\be
\left(\nabla_n+K_{\alpha,s}(q)\right)\phi_s=0,
\qquad
\nabla_n\equiv n_{(\alpha)}^\mu\partial_\mu.
\label{maxwell_bc}
\ee
The response is called \emph{linear} because the boundary equation is linear in the field amplitude, as follows from a quadratic surface action.

The corresponding reflection coefficient follows directly from Eq.~\eqref{maxwell_bc}. Let $y$ be a local coordinate increasing along the outward normal, with the plate at $y=0$, and write the local profile as $\phi_s=e^{-qy}+r_{\alpha,s}e^{qy}$. At the plate,
$\phi_s(0)=1+r_{\alpha,s}$ and $\partial_y\phi_s(0)=-q+qr_{\alpha,s}$. Inserting these two expressions into the Robin condition gives
\be
r_{\alpha,s}(q)=
\frac{q-K_{\alpha,s}(q)}{q+K_{\alpha,s}(q)}.
\label{maxwell_reflection}
\ee
This is the standard Robin reflection amplitude used in scattering formulations of the Casimir effect \cite{Lambrecht:2006few,Rahi:2009hm}. One complete round trip between the plates contains the product
\be
\rho_s(q)\equiv r_{0,s}(q)r_{h,s}(q)
\label{rho}
\ee
and the propagation factor $e^{-2qh}=e^{-2X}$.

The finite interaction energy is defined by the standard subtraction
\be
\Eint(h)\equiv E(h)-E(\infty).
\label{subtraction}
\ee
In a determinant formulation the unnormalized two-plate determinant contains three types of factors: a bulk normalization, a factor associated with each isolated plate, and a factor containing propagation from one plate to the other and back. The first two types remain when $h\to\infty$ and therefore cancel in Eq.~\eqref{subtraction}; only the two-surface interaction survives \cite{Milton:2001yy,Maggiore:2026agw,Blasi:1992mm}. Appendix~\ref{appendix_modesum} derives the same subtraction from the mode sum. For the two Maxwell polarizations the result is
\be
\frac{\Eint^{\rm EM}(h)}{\Acal}
=\frac12\sum_{s={\rm TE,TM}}\int\frac{\dd^3p}{(2\pi)^3}
\log\!\left[1-\rho_s(q)e^{-2X}\right].
\label{maxwell_energy_general}
\ee
Expanding the logarithm in powers of $\rho_s e^{-2X}$ makes the scattering interpretation explicit: the term of order $n$ corresponds to $n$ complete round trips between the plates.

For two perfectly conducting plates, TE and TM have opposite reflection signs at a single conductor, but each polarization is reflected by the same type of plate at both ends. Therefore the product is the same for both polarizations,
\be
\rho_{\rm TE}=\rho_{\rm TM}=1,
\label{perfect_products}
\ee
as in the standard ideal-conductor calculation \cite{Milton:2001yy,Bordag:2009zz}. Since the integrand then depends only on $q$, the three-dimensional momentum integral reduces to
\be
\frac12\int\frac{\dd^3p}{(2\pi)^3}f(qh)
=\frac{1}{4\pi^2h^3}\int_0^\infty\dd X\,X^2 f(X).
\label{radial_reduction}
\ee
For equal ideal reflection it is useful to denote the remaining radial integral by $I_+$:
\be
\begin{aligned}
I_+\equiv\int_0^\infty\dd X\,X^2\log\!\left(1-e^{-2X}\right)
&=-\sum_{n=1}^\infty\frac1n
\int_0^\infty\dd X\,X^2e^{-2nX}\\
&=-\frac14\sum_{n=1}^\infty\frac1{n^4}
=-\frac{\zeta(4)}4
=-\frac{\pi^4}{360},
\end{aligned}
\label{Iplus}
\ee
where $\zeta(s)=\sum_{n=1}^\infty n^{-s}$ is the Riemann zeta function.
For later comparison we also record the complementary ideal-reflection integral, obtained by expanding $\log(1+e^{-2X})$ in the same way,
\be
I_-\equiv\int_0^\infty\dd X\,X^2\log\!\left(1+e^{-2X}\right)
=\frac{7\pi^4}{2880}.
\label{Iminus}
\ee
For ordinary electromagnetism this corresponds to complementary ideal electric--magnetic reflection and gives the standard repulsive Boyer result \cite{Boyer:1974}.
 Combining Eqs.~\eqref{maxwell_energy_general}, \eqref{perfect_products}, \eqref{radial_reduction} and \eqref{Iplus} gives the standard Maxwell result
\be
\frac{E_{\rm EM}(h)}{\Acal}=-\frac{\pi^2}{720h^3}.
\label{intro_maxwell_energy}
\ee
The corresponding force is
\be
\frac{F_{\rm EM}(h)}{\Acal}
=-\frac{1}{\Acal}\frac{\partial E_{\rm EM}}{\partial h}
=-\frac{\pi^2}{240h^4},
\label{maxwell_force}
\ee
where the negative sign denotes attraction. This Maxwell calculation fixes both the normalization and the reflection language used for the higher-rank systems below.

\section{The full covariant higher-rank theory and BRS gauge fixing}
\label{section_tensor}

\subsection{Bulk action}

The scalar-charge gauge theory reviewed in the Introduction uses a symmetric spatial tensor because its Gauss law and gauge transformation contain two spatial derivatives. The Lorentz-covariant extension keeps the same idea while treating space and time on equal footing: the gauge potential is a symmetric rank-two spacetime tensor,
\be
A_{\mu\nu}=A_{\nu\mu},
\label{symtensor}
\ee
and its gauge transformation is the double-derivative scalar symmetry defined in Eq.~\eqref{intro_tensor_gauge}. This is the natural spacetime continuation of Eq.~\eqref{intro_scalar_charge_gauge}: the vector potential and one-derivative gauge transformation of ordinary Maxwell theory are replaced by a symmetric tensor potential and a two-derivative scalar transformation. This Maxwell-like covariant theory was introduced in Ref.~\cite{Bertolini:2022ijb}; a wider symmetric-tensor family was studied in Ref.~\cite{Blasi:2022mbl}.

The gauge-invariant rank-three field strength is
\be
F_{\mu\nu\rho}
=\partial_\mu A_{\nu\rho}+\partial_\nu A_{\mu\rho}-2\partial_\rho A_{\mu\nu},
\qquad
F_{\mu\nu\rho}=F_{\nu\mu\rho},
\label{tensor_field_strength}
\ee
and obeys the cyclic identity
\be
F_{\mu\nu\rho}+F_{\nu\rho\mu}+F_{\rho\mu\nu}=0.
\label{cyclic}
\ee
With the mostly-plus Lorentzian convention of Eq.~\eqref{metric}, the source-free action of Ref.~\cite{Bertolini:2022ijb} is
\be
S_{\rm F}^{\rm L}[A]=-\frac{g_{\rm F}}6\int_{\M}\dd^4x\,
F_{\mu\nu\rho}F^{\mu\nu\rho}.
\label{tensor_action_lorentzian}
\ee
After Wick rotation, the Euclidean action used in the functional determinant is
\be
S_{\rm F}[A]=\frac{g_{\rm F}}6\int_{\M}\dd^4x\,
F_{\mu\nu\rho}F_{\mu\nu\rho},
\label{tensor_action}
\ee
where $g_{\rm F}>0$ is an overall normalization of the free higher-rank action. Multiplying a quadratic operator by a constant changes its determinant by an overall factor that is the same at finite plate separation and at $h\to\infty$. The subtraction in Eq.~\eqref{subtraction} therefore removes the dependence on $g_{\rm F}$, just as it removes the normalization $g_{\rm EM}$ in Maxwell theory.

Equation~\eqref{tensor_action} is a sum of Euclidean squares, so for $g_{\rm F}>0$ its gauge-invariant quadratic form has no negative directions. It does have zero directions: a field variation of the form $\delta A_{\mu\nu}=\partial_\mu\partial_\nu\Lambda$ leaves the field strength unchanged and therefore costs no action. These are the gauge zero modes. BRS gauge fixing removes this degeneracy and makes the gauge-fixed quadratic operator non-degenerate on the boundary domain used in the slab calculation. A positive Euclidean quadratic form is not, however, the same statement as a stable real-time theory. Lorentzian stability requires a Hamiltonian bounded from below; Section~\ref{section_parent_vector_stability} checks that condition directly for the full covariant theory.

The Euler--Lagrange equations obtained by varying Eq.~\eqref{tensor_action} are
\be
M_{\mu\nu}\equiv\partial^\rho F_{\mu\nu\rho}
=\partial_\mu V_\nu+\partial_\nu V_\mu-2\Box A_{\mu\nu}=0,
\qquad
V_\nu=\partial^\rho A_{\nu\rho},
\qquad
\Box=\partial^\rho\partial_\rho.
\label{tensor_eom}
\ee
The equations satisfy the Noether identity
\be
\partial^\mu\partial^\nu M_{\mu\nu}\equiv0,
\label{noether_identity}
\ee
which follows from the scalar gauge symmetry.

\subsection{Covariant BRS gauge fixing}
\label{section_brs_gauge_fixing}

The quadratic operator following from Eq.~\eqref{tensor_action} has a null direction generated by Eq.~\eqref{intro_tensor_gauge}. A gauge condition is therefore required to define the propagator. Physical observables must be independent of that choice. This is the main role of the BRS formulation in the present calculation: changes of the covariant gauge condition enter the complete action through BRS-exact terms, which is the situation covered by the algebraic gauge-independence analysis of Piguet and Sibold \cite{Piguet:1984js}. Section~\ref{section_brs_boundary} carries the argument out on the slab, where the boundary domain has to be BRS closed as well.

If the gauge condition is required to be local, Lorentz scalar, linear in $A_{\mu\nu}$ and to respect the power counting, only two independent scalars are available: $\partial^\mu\partial^\nu A_{\mu\nu}$ and $\Box A$. Up to an overall normalization, the most general gauge function is therefore
\be
G_\chi[A]
=\partial^\mu\partial^\nu A_{\mu\nu}+\chi\,\Box A,
\qquad
A\equiv A^\mu{}_{\mu},
\label{covariant_gauge_family}
\ee
This scalar covariant gauge family is the one discussed for the same higher-rank symmetry in Refs.~\cite{Bertolini:2023gauging,Fecit:Rovere:2026worldline}. It transforms as
\be
\delta G_\chi=(1+\chi)\Box^2\Lambda.
\label{general_gauge_variation}
\ee
At $\chi=-1$, the function $G_{-1}$ is gauge invariant and leaves the scalar gauge freedom unfixed. Every fixed $\chi\neq-1$ therefore defines a valid covariant gauge function. The parameter $\chi$ labels the choice of gauge condition within this family. The usual gauge parameter $\xi>0$ controls the weight with which the chosen condition is imposed in the action.

We introduce a ghost $c$, an antighost $\bar c$ and the Lagrange multiplier $B$, with BRS transformations \cite{Becchi:1975nq,Becchi:1976nq,Baulieu:1985jc}
\be
sA_{\mu\nu}=\partial_\mu\partial_\nu c,
\qquad
sc=0,
\qquad
s\bar c=B,
\qquad
sB=0.
\label{brs_transformations}
\ee
The differential $s$ is nilpotent off shell, that is, $s^2=0$ without using any equation of motion. The gauge-fixing and ghost terms are produced together as the BRS variation of a single anticommuting functional. We use the standard term \emph{gauge-fixing fermion}; this terminology is particularly natural in the Batalin--Vilkovisky formulation \cite{Batalin:1981jr,Batalin:1983gg}. The bulk gauge-fixing fermion is
\be
\Psi_{\xi,\chi}^{\rm bulk}
=g_{\rm F}\int_{\M}\dd^4x\,
\bar c\left(G_\chi[A]-\frac{\xi}{2}B\right),
\label{bulk_gauge_fermion}
\ee
and its BRS variation gives
\be
\begin{aligned}
S_{\rm gf+FP}
=s\Psi_{\xi,\chi}^{\rm bulk}
=g_{\rm F}\int_{\M}\dd^4x\,
\left[
B G_\chi[A]-\frac{\xi}{2}B^2
-(1+\chi)\bar c\Box^2c
\right].
\end{aligned}
\label{brs_gauge_action}
\ee
The algebraic equation of the multiplier is
\be
B=\frac{1}{\xi}G_\chi[A].
\label{B_equation}
\ee
Eliminating the auxiliary multiplier $B$ gives the usual second-order gauge-fixing term
\be
S_{\rm gf}
=\frac{g_{\rm F}}{2\xi}\int_{\M}\dd^4x\,G_\chi[A]^2,
\label{gauge_fixing_action}
\ee
which is manifestly positive for $\xi>0$. Since $G_\chi$ contains two derivatives, $\xi$ has mass dimension two. The Faddeev--Popov operator is
\be
\mathcal M_{{\rm FP},\chi}=(1+\chi)\Box^2.
\label{FP_operator}
\ee
The Abelian gauge symmetry makes this operator field independent. On a slab, however, its spectrum depends on the plate separation once the boundary domain is specified \cite{FaddeevPopov:1967fc}.

\subsection{BRS-compatible boundary completion and gauge-parameter independence}
\label{section_brs_boundary}

The boundary problem has two logically distinct parts. The physical response of the plates is discussed later in Section~\ref{section_boundary_doublets}, where it is derived from a local surface action. The present subsection fixes instead the gauge-dependent sector and the Faddeev--Popov domain.

At each plate define the local tangential gauge functions
\be
\mathcal F_{1,\alpha}
=\left.\partial^a\partial^b A_{ab}\right|_{\Sigma_\alpha},
\qquad
\mathcal F_{2,\alpha}
=\left.\partial^a A_{a3}\right|_{\Sigma_\alpha}.
\label{boundary_gauge_functions}
\ee
Introduce two boundary BRS doublets \cite{PiguetSorella},
\be
s\bar\eta_\alpha=b_\alpha,
\qquad
sb_\alpha=0,
\qquad
s\bar\vartheta_\alpha=d_\alpha,
\qquad
sd_\alpha=0,
\label{boundary_brs_doublets}
\ee
A BRS doublet is cohomologically trivial: its two members form an exact pair and do not change the BRS cohomology, hence the physical observables \cite{PiguetSorella}. We use these doublets to define the boundary gauge sector through the local boundary gauge-fixing fermion
\be
\Psi_\partial
=\sum_{\alpha=0,h}\int_{\Sigma_\alpha}\dd^3x\,
\left(\bar\eta_\alpha\mathcal F_{1,\alpha}
+\bar\vartheta_\alpha\mathcal F_{2,\alpha}\right).
\label{boundary_gauge_fermion}
\ee
Its BRS variation is
\be
s\Psi_\partial
=\sum_{\alpha=0,h}\int_{\Sigma_\alpha}\dd^3x\,
\left[
 b_\alpha\mathcal F_{1,\alpha}
+d_\alpha\mathcal F_{2,\alpha}
-\bar\eta_\alpha(\partial_\parallel^2)^2c
-\bar\vartheta_\alpha\partial_\parallel^2\partial_3c
\right],
\label{boundary_brs_action}
\ee
where $\partial_\parallel^2\equiv\partial^a\partial_a$. The boundary multipliers $b_\alpha$ and $d_\alpha$ impose $\mathcal F_{1,\alpha}=\mathcal F_{2,\alpha}=0$. For a fixed nonzero tangential momentum,
\be
\mathcal F_{1,\alpha}=-q^2A_{pp},
\qquad
\mathcal F_{2,\alpha}=iqA_{p3},
\label{boundary_gauge_functions_fourier}
\ee
so these equations are equivalent to
\be
A_{pp}=0,
\qquad
A_{p3}=0
\qquad
\text{on }\Sigma_0\text{ and }\Sigma_h.
\label{longitudinal_essential_bc}
\ee
The equations of motion of the boundary antighosts impose $(\partial_\parallel^2)^2c=0$ and $\partial_\parallel^2\partial_3c=0$. For $q>0$ this gives the clamped ghost domain
\be
c=0,
\qquad
\partial_3c=0
\qquad
\text{on }\Sigma_0\text{ and }\Sigma_h.
\label{FP_clamped_domain}
\ee
Here \emph{clamped} is used in the sense standard for fourth-order problems: the field and its first normal derivative vanish at each plate. These are two conditions per boundary, the number that a fourth-order normal operator requires, exactly as for a clamped elastic plate. On this domain, $c=\partial_3c=0$, the BRS variations of the tangential tensor components that enter the physical surface action vanish at each plate. The physical surface functional discussed in Section~\ref{section_boundary_doublets} is therefore BRS invariant on the same BRS-closed domain, even though its individual terms need not be gauge invariant away from the boundary.

The same local BRS-exact boundary functional therefore generates the bosonic boundary gauge slice and the corresponding ghost domain. The decomposition of the gauge-invariant tensor components and their physical surface response are derived separately in Sections~\ref{section_modes} and \ref{section_boundary_doublets}.

The fourth-order ghost operator is paired with the same clamped adjoint domain,
\be
\bar c=0,
\qquad
\partial_3\bar c=0
\qquad
\text{on }\Sigma_0\text{ and }\Sigma_h.
\label{antighost_clamped_domain}
\ee
This choice makes the Green boundary form of $\Box^2$ vanish, that is, the surface term that survives when the fourth-order operator is moved from $c$ to $\bar c$ by repeated integration by parts. Its vanishing is what makes the ghost operator self-adjoint on the domain chosen above. Off-shell BRS closure then requires
\be
B=0,
\qquad
\partial_3B=0
\qquad
\text{on }\Sigma_0\text{ and }\Sigma_h,
\label{B_clamped_domain}
\ee
because $s\bar c=B$. After the multiplier is eliminated through Eq.~\eqref{B_equation}, the multiplier domain in Eq.~\eqref{B_clamped_domain} becomes
\be
G_\chi=0,
\qquad
\partial_3G_\chi=0
\qquad
\text{on }\Sigma_0\text{ and }\Sigma_h.
\label{longitudinal_natural_bc}
\ee
The first-order BRS formulation thus maps the clamped multiplier domain into the natural conditions on $G_\chi$. Appendix~\ref{appendix_variational_gauge} starts from the same essential boundary gauge slice and derives Eq.~\eqref{longitudinal_natural_bc} directly from the classical variational principle, without ghosts or multiplier. Appendix~\ref{appendix_gauge_determinant} then verifies the off-shell BRS closure and the longitudinal--Faddeev--Popov cancellation. The BRS route proves gauge-parameter independence algebraically for arbitrary $\xi$ and $\chi\neq-1$, while the variational route reproduces the same natural boundary conditions throughout that gauge family. The explicit determinant cancellation is checked separately at $\chi=0$ with arbitrary $\xi>0$, where the transverse trace and longitudinal block decouple. These statements provide complementary, but deliberately distinct, checks of the gauge-sector result.

The total gauge-fixed action is
\be
S_{\rm tot}(\xi,\chi)
=S_{\rm F}+S_{\partial}^{\rm phys}
+s\left(\Psi_{\xi,\chi}^{\rm bulk}+\Psi_\partial\right).
\label{total_brs_action}
\ee
The normalized Euclidean interaction functional is gauge invariant, and the BRS formulation expresses its gauge-parameter independence through the BRS-exact variations of the complete action, following the algebraic construction of Ref.~\cite{Piguet:1984js}:
\be
\frac{\partial S_{\rm tot}}{\partial\xi}
=s\left[-\frac{g_{\rm F}}2\int_{\M}\dd^4x\,\bar c B\right],
\qquad
\frac{\partial S_{\rm tot}}{\partial\chi}
=s\left[g_{\rm F}\int_{\M}\dd^4x\,\bar c\Box A\right].
\label{gauge_parameter_exactness}
\ee
Let $Z(h;\xi,\chi)$ denote the source-free Euclidean functional integral on the slab, and define $\Gamma(h;\xi,\chi)\equiv-\log Z(h;\xi,\chi)$. For the unrestricted covariant theory, its subtracted finite part is the normalized Euclidean interaction functional analyzed below; no assumption of a stable Lorentzian vacuum is needed for the BRS identity. The following identities apply to this vacuum functional. A generic off-shell one-particle-irreducible effective action may depend on gauge parameters; this does not contradict the gauge independence of the source-free vacuum functional considered here. The functional measure of the present free Abelian theory is BRS invariant, and the boundary domain defined above is mapped into itself by Eq.~\eqref{brs_transformations}. A BRS change of variables therefore gives
\be
\frac{\partial\Gamma(h;\xi,\chi)}{\partial\xi}=0,
\qquad
\frac{\partial\Gamma(h;\xi,\chi)}{\partial\chi}=0.
\label{gauge_parameter_independence_full}
\ee
The same identities hold for the finite interaction functional
\be
\Gamma_{\rm int}(h)=\Gamma(h)-\Gamma(\infty).
\label{interaction_effective_action}
\ee
Equations~\eqref{gauge_parameter_independence_full} establish gauge-parameter independence before evaluating the longitudinal and Faddeev--Popov determinants. The spectral calculation can therefore use $\chi=0$, where the gauge-invariant transverse trace separates from the longitudinal block, while $\xi$ remains arbitrary. Section~\ref{section_gauge_determinant} displays the cancellation in this gauge. The explicit spectral cancellation below is evaluated at $\chi=0$ with arbitrary $\xi>0$. Independence of the complete normalized functional from $\chi$ follows from the BRS identity in Eq.~\eqref{gauge_parameter_independence_full}.

The parallel between the two theories is now explicit:
\begin{center}
\begin{tabular}{@{}>{\raggedright\arraybackslash}p{0.23\textwidth}>{\raggedright\arraybackslash}p{0.28\textwidth}>{\raggedright\arraybackslash}p{0.35\textwidth}@{}}
\toprule
 & Maxwell theory & full higher-rank theory \\
\midrule
gauge potential & $A_\mu$ & $A_{\mu\nu}=A_{\nu\mu}$ \\
scalar gauge transformation & $\delta A_\mu=\partial_\mu\Lambda$ & $\delta A_{\mu\nu}=\partial_\mu\partial_\nu\Lambda$ \\
gauge-invariant curvature & $F_{\mu\nu}$ & $F_{\mu\nu\rho}$ \\
gauge-invariant normal sectors & two independent channels & three independent Maxwell-like modes and two coupled vector pairs \\
\bottomrule
\end{tabular}
\end{center}
Section~\ref{section_modes} derives the last row explicitly. The coupled vector pairs will then be compared with the stable radiative content of the fractonic branch.

\section{Normal-mode decomposition across the slab}
\label{section_modes}

\subsection{Fourier transform and polarization basis}

We Fourier transform the full covariant tensor field with respect to the coordinates parallel to the plates and write the transformed field as $\widetilde A_{\mu\nu}(p,x^3)$. The only coordinate left untransformed is the normal coordinate $x^3$. To determine its possible dependence on the normal coordinate, we use
\be
\widetilde A_{\mu\nu}(p,x^3)
=a_{\mu\nu}(p)e^{\kappa x^3}.
\label{normal_ansatz}
\ee
The bulk equations determine the allowed values of $\kappa$, while the normalized two-plate interaction is obtained by integrating the corresponding finite determinant over the tangential momentum $p_a$.

For the bulk decomposition, let $n^\mu=\delta^\mu_{\ 3}$ denote the fixed unit vector along increasing $x^3$; the outward plate normals in Eq.~\eqref{normals} are $n^{(0)}=-n$ and $n^{(h)}=n$. For a fixed nonzero $p_a$, let $\widehat p^\mu$ be the unit vector in its direction and let $e_{(1)}^\mu$ and $e_{(2)}^\mu$ be orthonormal four-vectors tangent to the plates and orthogonal to $\widehat p^\mu$. The circular vectors
\be
m_\pm^\mu
=\frac{e_{(1)}^\mu\pm i e_{(2)}^\mu}{\sqrt2}
\label{circular_vectors}
\ee
acquire phases $e^{\pm i\varphi}$ under rotations in the transverse plane. These rotations leave $p_a$ fixed and form its $SO(2)$ stabilizer, namely the subgroup of tangential rotations that preserves the chosen momentum. The label $s$ below denotes the corresponding $SO(2)$ weight.

From this point on, component labels refer to contractions of the Fourier-transformed field $\widetilde A_{\mu\nu}(p,x^3)$; the tilde is suppressed on those labels. In particular,
\be
\begin{aligned}
A_{m^\pm p}&\equiv m_\pm^\mu\widehat p^\nu\widetilde A_{\mu\nu},
&
A_{m^\pm3}&\equiv m_\pm^\mu n^\nu\widetilde A_{\mu\nu},\\
A_{pp}&\equiv\widehat p^\mu\widehat p^\nu\widetilde A_{\mu\nu},
&
A_{p3}&\equiv\widehat p^\mu n^\nu\widetilde A_{\mu\nu},\\
A_{33}&\equiv n^\mu n^\nu\widetilde A_{\mu\nu},
&
A_{m^+m^-}&\equiv m_+^\mu m_-^\nu\widetilde A_{\mu\nu}.
\end{aligned}
\label{component_contractions}
\ee
The ten components of the symmetric tensor then separate into independent blocks:
\begin{center}
\begin{tabular}{@{}lll@{}}
\toprule
$SO(2)$ weight & components & role in the full-theory determinant \\
\midrule
$s=\pm2$ & $A_{m^\pm m^\pm}$ & independent \\
$s=\pm1$ & $u_\pm=A_{m^\pm p}$, $v_\pm=A_{m^\pm3}$ & coupled pairs \\
$s=0$ & $t=A_{m^+m^-}$ & independent \\
$s=0$ & $(A_{pp},A_{p3},A_{33})$ & gauge-dependent block \\
\bottomrule
\end{tabular}
\end{center}
The seven components listed in the first three rows are gauge invariant under Eq.~\eqref{intro_tensor_gauge}. The three components in the last row contain the direction generated by the generalized gauge transformation and are controlled by the covariant gauge choice of Section~\ref{section_tensor}. The tangential Fourier conventions and the algebra of the bulk symbol that lead to this decomposition are collected in Appendix~\ref{appendix_modes}.

\subsection{Three independent Maxwell-like modes}

For $A_{m^+m^+}$, $A_{m^-m^-}$ and $t$, the field equations reduce to
\be
\left(-\partial_3^2+q^2\right)\phi=0,
\label{tensor_helmholtz}
\ee
where $\phi$ denotes any one of these three amplitudes. The two $s=\pm2$ components and the transverse trace $t$ are therefore three independent Maxwell-like modes. Their propagation across the slab is identical to that of a Maxwell polarization, and their contribution to the normalized full-theory interaction follows from the one-channel factor in Eq.~\eqref{maxwell_energy_general}.

\subsection{Two coupled Jordan doublets}

The $s=\pm1$ components remain coupled. For either value of $s$, write the pair as
\be
\Psi_s=
\begin{pmatrix}u_s\\v_s\end{pmatrix}.
\label{doublet_vector}
\ee
Suppressing the label $s$ when no ambiguity arises, the normal equations are
\be
-2u''+q^2u+iqv'=0,
\qquad
iqu'-v''+2q^2v=0,
\label{doublet_equations}
\ee
where a prime denotes $\partial_3$. Substituting Eq.~\eqref{normal_ansatz} and writing the two amplitudes as the column vector $\mathbf a=(a_u,a_v)^T$ gives $\mathcal M_1(\kappa)\mathbf a=0$, with
\be
\mathcal M_1(\kappa)=
\begin{pmatrix}
-2\kappa^2+q^2 & iq\kappa\\
iq\kappa & -\kappa^2+2q^2
\end{pmatrix},
\qquad
\det\mathcal M_1=2(\kappa^2-q^2)^2.
\label{doublet_matrix}
\ee
The characteristic roots are $\kappa=\pm q$, each with algebraic multiplicity two. Since $q>0$ in the Euclidean momentum integral, these roots are real. At either root, however, the kernel of $\mathcal M_1$ is one-dimensional. The geometric multiplicity is therefore smaller than the algebraic multiplicity. Such a repeated root is called defective \cite{HornJohnson,GohbergLancasterRodman,Kato}.

The missing independent solutions are generalized eigenvectors. A complete basis is
\be
\begin{pmatrix}u\\v\end{pmatrix}
=\begin{pmatrix}\pm i\\1\end{pmatrix}e^{\pm qx^3},
\qquad
\begin{pmatrix}u\\v\end{pmatrix}
=\begin{pmatrix}\pm ix^3\\x^3\pm3/q\end{pmatrix}e^{\pm qx^3},
\label{doublet_basis}
\ee
with correlated upper and lower signs. The first profile at each root is an ordinary eigenvector; the second is its generalized partner. Together they form a Jordan chain. Each gauge-invariant pair $\Psi_+$ or $\Psi_-$ is therefore a Jordan doublet.

This structure distinguishes the normal propagation from that of Maxwell theory. A photon polarization has one growing and one decaying exponential. A Jordan doublet has four independent profiles, two of which contain the additional factor $x^3$. When conditions are imposed at both plates, the polynomial factor survives in the boundary determinant. The full-theory interaction determinant therefore resolves the coupled vector propagation in addition to the multiplicity of gauge-invariant components.

\subsection{Gauge-dependent block and explicit BRS cancellation}
\label{section_gauge_determinant}

The remaining components
\be
\Phi=
\begin{pmatrix}A_{pp}\\ A_{p3}\\ A_{33}\end{pmatrix}
\label{longitudinal_vector}
\ee
form a three-component gauge-dependent block. For a generic value of the gauge parameter $\chi$ defined in Eq.~\eqref{covariant_gauge_family}, the off-shell gauge-fixing term also mixes this block with the transverse trace $t=A_{m^+m^-}$ defined in Eq.~\eqref{component_contractions}. The reason is direct: the gauge function $G_\chi$ contains the spacetime trace $A$, and that trace contains $t$ together with the longitudinal components. The choice $\chi=0$ removes this mixing. The independence established in Eq.~\eqref{gauge_parameter_independence_full}, and independently checked in Appendix~\ref{appendix_variational_gauge}, allows the explicit determinant calculation to be performed in the convenient gauge
\be
\chi=0,
\qquad
G[A]=\partial^\mu\partial^\nu A_{\mu\nu},
\qquad
\mathcal M_{\rm FP}=\Box^2,
\label{convenient_gauge}
\ee
without restricting the physical result. We keep $\xi$ arbitrary.

On both plates, the BRS boundary gauge-fixing fermion \eqref{boundary_gauge_fermion} gives $A_{pp}=A_{p3}=0$ and the clamped ghost domain $c=c'=0$. The antighost obeys the adjoint clamped conditions on both plates, and BRS closure gives $B=B'=0$ there. Eliminating the multiplier through Eq.~\eqref{B_equation} maps the latter conditions into $G=G'=0$ in the second-order formulation.

Let $N_\xi$ denote the resulting gauge-fixed operator acting on the three-component vector $\Phi$ defined in Eq.~\eqref{longitudinal_vector}. For fixed $q>0$, the solution space is eight-dimensional. It separates into two fourth-order scalar structures: one generated by the gauge parameter $\Lambda$ and one by the gauge function $G$. Each fourth-order equation supplies four independent normal profiles. The first four solutions are pure-gauge configurations
\be
\Phi_{\Lambda}
=
\begin{pmatrix}
-q^2\Lambda\\ iq\Lambda'\\ \Lambda''
\end{pmatrix},
\qquad
\left(\partial_3^2-q^2\right)^2\Lambda=0.
\label{pure_gauge_family}
\ee
The remaining four solutions are distinguished by a nonzero value of the gauge function $G$ defined in Eq.~\eqref{convenient_gauge}. Applying the adjoint of the gauge generator defined by Eq.~\eqref{intro_tensor_gauge} to the gauge-fixed equations derived from Eqs.~\eqref{tensor_eom} and \eqref{gauge_fixing_action}, and then using the Noether identity \eqref{noether_identity}, gives
\be
\left(\partial_3^2-q^2\right)^2G=0.
\label{gauge_function_equation}
\ee
The longitudinal solution space therefore contains two four-dimensional copies of the same clamped fourth-order scalar problem: one parametrized by $\Lambda$ and one parametrized by $G$. For any linear boundary problem, imposing all boundary conditions on a basis of bulk solutions gives a homogeneous matrix equation for the expansion coefficients. We call this matrix the boundary, or secular, matrix; its determinant vanishes exactly when a nonzero solution satisfies all boundary conditions. Appendix~\ref{appendix_gauge_determinant} proves the factorization above and constructs a basis in which the resulting $8\times8$ boundary matrix is block triangular.

Denote by $D_{\rm FP}(q,h)$ the secular determinant of the clamped operator $(\partial_3^2-q^2)^2$. The longitudinal determinant factorizes as
\be
D_N(q,h;\xi)
=C(q,\xi)\,D_{\rm FP}^2(q,h),
\label{unnormalized_longitudinal_factorization}
\ee
where $C(q,\xi)$ contains only basis-normalization and gauge-parameter factors. These factors are local to the fixed-$q$ solution basis and do not contain propagation between the two plates; hence they are independent of the separation $h$. Appendix~\ref{appendix_gauge_determinant} gives the explicit factorization. After division by the corresponding $h\to\infty$ factors,
\be
\Delta_N(X)=\Delta_{\rm FP}^2(X),
\label{normalized_longitudinal_factorization}
\ee
where $X$ is the dimensionless variable defined in Eq.~\eqref{tangential_momentum}.
For the clamped scalar problem, the four normal profiles are
\be
 e^{qx^3},
\qquad
 x^3e^{qx^3},
\qquad
 e^{-qx^3},
\qquad
 x^3e^{-qx^3},
\label{FP_normal_basis}
\ee
and the normalized interaction factor is
\be
\Delta_{\rm FP}(X)
=1-\left(2+4X^2\right)e^{-2X}+e^{-4X}.
\label{FP_interaction_factor}
\ee
The real bosonic block contributes $\frac12\log\Delta_N$, while the complex ghost contributes $-\log\Delta_{\rm FP}$. Hence
\be
\frac12\log\Delta_N(X)-\log\Delta_{\rm FP}(X)=0.
\label{FP_longitudinal_cancellation}
\ee
The BRS identities in Eq.~\eqref{gauge_parameter_independence_full} prove that this cancellation is independent of $\xi$ and $\chi$; Eq.~\eqref{FP_longitudinal_cancellation} displays it explicitly in the gauge \eqref{convenient_gauge}. The parameter independence with respect to $\chi$ is established by the BRS identity. The explicit determinant factorization is displayed in the decoupling gauge $\chi=0$. The gauge-dependent sector therefore contributes no plate-separation dependence. The normalized determinant of the full covariant theory is therefore carried by the three independent Maxwell-like modes and the two coupled Jordan doublets.

The helicity decomposition and the boundary gauge functions degenerate at the isolated point $q=0$. That zero-momentum gauge sector is divided by its residual gauge volume before the momentum integral is formed. It is a single point in the three-dimensional integral and does not change the energy per unit area.

\subsection{Lorentzian status of the Jordan sector}
\label{section_parent_vector_stability}

The physical meaning of the Jordan doublets is determined by their real-time dynamics. We therefore analyze the same two-component sector using Lorentzian time as the evolution coordinate before assigning a vacuum-energy interpretation to the Euclidean determinant.

Choose a Lorentzian plane wave with spatial momentum along $x^3$ and consider one transverse-vector sector,
\be
u=A_{13},
\qquad
v=A_{01}.
\label{parent_vector_fields}
\ee
The opposite helicity gives the same equations because the two $SO(2)$ weights $s=+1$ and $s=-1$ are related by complex conjugation; it therefore has the same spectrum and Hamiltonian structure. Substituting these two fields into the Lorentzian full-theory action \eqref{tensor_action_lorentzian} gives, up to the common factor $g_{\rm F}$,
\be
\mathcal L_V
=2\dot u^{\,2}-2\dot u\,\partial_3v-(\partial_3u)^2
-\dot v^{\,2}+2(\partial_3v)^2.
\label{parent_vector_lagrangian}
\ee
The velocity Hessian is
\be
\frac{\partial^2(\mathcal L_V/g_{\rm F})}
{\partial(\dot u,\dot v)^2}
=\begin{pmatrix}4&0\\0&-2\end{pmatrix},
\qquad
\det=-8.
\label{parent_vector_hessian}
\ee
Thus both variables are canonical and their kinetic quadratic form is indefinite. With momenta $p_u$ and $p_v$, the Hamiltonian density is
\be
\frac{\mathcal H_V}{g_{\rm F}}
=\frac{(p_u+2\partial_3v)^2}{8}
-\frac{p_v^2}{4}
+(\partial_3u)^2-2(\partial_3v)^2,
\label{parent_vector_hamiltonian}
\ee
which is not bounded below.

The Euler--Lagrange equations are
\be
2\ddot u-u''-\dot v'=0,
\qquad
\ddot v-2v''+\dot u'=0.
\label{parent_vector_eom}
\ee
For a plane wave $e^{-i\omega t+ikx^3}$, replacing $\partial_t$ by $-i\omega$ and $\partial_3$ by $ik$ turns the differential equations into a $2\times2$ algebraic matrix. This matrix is the Fourier symbol of the vector equations, and its determinant is proportional to
\be
(\omega^2-k^2)^2.
\label{parent_vector_temporal_det}
\ee
At $\omega=\pm k$ the symbol has rank one, so each repeated root has only one ordinary eigenvector. Generalized solutions proportional to $t e^{\mp ikt}$ are therefore present. The Jordan structure of the Euclidean normal problem is consequently a property of the transverse-vector dynamics of the full theory, not an artifact of choosing $x^3$ as the normal evolution coordinate.

The longitudinal-diffeomorphism fracton-gravity framework, including its additional lower-spin sectors, was developed in Ref.~\cite{Afxonidis:2024spacetime}; a spin-one instability and an unbounded Hamiltonian were subsequently identified in that broader class \cite{Afxonidis:2024tph}. Here Eqs.~\eqref{parent_vector_lagrangian}--\eqref{parent_vector_temporal_det} establish the result directly for the specific Maxwell-like $F_{\mu\nu\rho}F^{\mu\nu\rho}$ theory used in this paper.

This changes the interpretation, but not the algebra, of the Euclidean slab calculation. The three one-component blocks and the two two-component Jordan blocks contain seven gauge-invariant Gaussian components. Their normalized Euclidean determinants are well-defined spectral quantities. Because the Lorentzian vector sector has no Hamiltonian bounded below, however, we shall refer to the result obtained from all seven components as the \emph{unrestricted covariant interaction determinant}, rather than as the vacuum energy of a stable fractonic phase. The stable fractonic branch is constructed separately in Section~\ref{section_fractonic_branch}.

\section{Surface response of the full covariant theory and the determinant of a Jordan doublet}
\label{section_boundary_doublets}

The bulk equations determine the profiles available in the slab. The plate action determines which linear combinations are allowed at each surface. For an independent Maxwell-like mode this information is summarized by a scalar reflection coefficient. For a Jordan doublet, the two components remain coupled, so the boundary response is a $2\times2$ matrix.

\subsection{Boundary variation and the local surface class}

Varying the bulk action \eqref{tensor_action} and integrating by parts gives the field equations in the interior and a surface contribution. We use the orientation sign $\sigma_0=-1$ on the lower plate and $\sigma_h=+1$ on the upper plate, so that
\be
\delta S_{\rm F}\big|_{\partial\M}
=2g_{\rm F}\sum_{\alpha=0,h}\sigma_\alpha\int_{\Sigma_\alpha}\dd^3x\,
\left[\pi^{ab}\delta A_{ab}+\pi^a\delta A_{a3}\right],
\label{boundary_variation}
\ee
where
\be
\pi_{ab}=\partial_3A_{ab}-\partial_{(a}A_{b)3},
\qquad
\pi_a=\partial_3A_{a3}-\partial_aA_{33}.
\label{boundary_momenta}
\ee
Parentheses denote symmetrization with unit weight. The quantities $\pi_{ab}$ and $\pi_a$ are the momenta conjugate to the boundary values of $A_{ab}$ and $A_{a3}$ with respect to evolution along increasing $x^3$; the outward-normal sign is the explicit factor $\sigma_\alpha$ in Eq.~\eqref{boundary_variation}. Without a surface action, stationarity of the bulk action sets the corresponding boundary variation to zero. A surface action changes these natural conditions and describes the response of the plates.

We consider a minimal local Hermitian surface class: the action is quadratic in the boundary fields, contains no normal derivatives and contains at most one tangential derivative. Terms without derivatives produce ordinary Robin response. A single tangential derivative is sufficient to include orientation-dependent interactions. Excluding normal derivatives keeps the order of the variational problem unchanged and avoids introducing additional boundary data. Higher tangential derivatives describe more general dispersive responses beyond the representative boundary conditions studied here. A tensor form of this local class is given in Appendix~\ref{appendix_surface}.

For one doublet, with $\Psi_s=(u_s,v_s)^T$ defined in Eq.~\eqref{doublet_vector}, define the boundary response matrix
\be
\mathcal K_{\alpha,s}(q)=
\begin{pmatrix}
a_{\alpha,s}(q) & b_{\alpha,s}(q)\\
\bar b_{\alpha,s}(q) & c_{\alpha,s}(q)
\end{pmatrix},
\label{response_matrix}
\ee
and the corresponding surface action
\be
S_{\partial,\alpha}^{(s)}
=\frac{g_{\rm F}}2\int\frac{\dd^3p}{(2\pi)^3}\,
\Psi_s^\dagger\mathcal K_{\alpha,s}\Psi_s.
\label{projected_surface_action}
\ee
The coefficient $a_{\alpha,s}$ multiplies $|u_s|^2$, $c_{\alpha,s}$ multiplies $|v_s|^2$, and $b_{\alpha,s}$ couples the two components. Hermiticity requires the lower off-diagonal entry to be the complex conjugate $\bar b_{\alpha,s}$.

Let
\be
S_{\rm tot}=S_{\rm F}+\sum_{\alpha=0,h}\sum_{s=\pm1}S_{\partial,\alpha}^{(s)}
\label{total_action_doublet}
\ee
for the doublet sector. Combining the surface term from the bulk variation with the variation of Eq.~\eqref{projected_surface_action} gives
\be
\sigma_\alpha(2u_s'-iqv_s)+a_{\alpha,s}u_s+b_{\alpha,s}v_s=0,
\qquad
\sigma_\alpha v_s'+\bar b_{\alpha,s}u_s+c_{\alpha,s}v_s=0,
\label{doublet_bc}
\ee
where $\sigma_0=-1$ and $\sigma_h=+1$. The term $-iqv_s$ in the first equation is present even when every surface coupling vanishes. It originates from the bulk momentum in Eq.~\eqref{boundary_momenta}. Their coupling is already encoded in the bulk variational problem and remains present when the physical plate action is set to zero.

\subsection{Boundary determinant and normalized interaction functional}

Expand a solution as
\be
\Psi_s(x^3)=\sum_{r=1}^{4}C_r\,\varphi_r(x^3),
\label{doublet_expansion}
\ee
where $\varphi_r$ are the four profiles in Eq.~\eqref{doublet_basis}. The two equations in Eq.~\eqref{doublet_bc} at each plate give four homogeneous linear equations for the four coefficients $C_r$. Using the definition of a boundary (secular) matrix given in Section~\ref{section_gauge_determinant}, we write them as
\be
\mathcal S_s(q,h)\,C=0.
\label{secular_system}
\ee
A nonzero mode exists only when the corresponding boundary determinant vanishes,
\be
D_s(q,h)\equiv\det\mathcal S_s(q,h)=0.
\label{secular_determinant}
\ee
The factors in $D_s$ that survive as $h\to\infty$ describe the two isolated plates. Dividing them out defines the finite interaction factor
\be
\Delta_{1,s}(q,h)
=\frac{D_s(q,h)}{D_s(q,\infty)}\Bigg|_{\rm finite},
\qquad
\Delta_{1,s}(q,h)\longrightarrow1
\quad(h\to\infty).
\label{doublet_interaction_factor}
\ee
The notation means that exponentially growing basis normalizations and factors associated with an isolated plate are divided out. They are independent of the interaction between the two surfaces and cancel in the subtraction at $h\to\infty$; Appendix~\ref{appendix_surface} displays the cancellation explicitly.

A real bosonic Gaussian integral contributes one half of the logarithm of its determinant. The normalized interaction functional of one doublet is therefore
\be
\frac{\mathcal E_{1,s}(h)}{\Acal}
=\frac12\int\frac{\dd^3p}{(2\pi)^3}\,
\log\Delta_{1,s}(q,h).
\label{doublet_energy_general}
\ee
This is the two-component analogue of the one-channel formula in Eq.~\eqref{maxwell_energy_general}.

\subsection{Three representative boundary conditions}

We now consider three choices that make the comparison with Maxwell transparent.

\subsubsection{Dirichlet boundary conditions}
Both components are fixed to zero on both plates. They can be obtained as the strong-coupling limit of positive diagonal surface terms in Eq.~\eqref{projected_surface_action}:
\be
u_s(0)=v_s(0)=u_s(h)=v_s(h)=0.
\label{dirichlet_bc}
\ee
Substituting the four profiles in Eq.~\eqref{doublet_basis} into these conditions gives a $4\times4$ boundary matrix. Its determinant is
\be
D_{\rm D}(q,h)=C_{\rm D}(q)\left[9\sinh^2X-X^2\right],
\label{dirichlet_secular}
\ee
where $C_{\rm D}(q)$ is independent of the plate separation. After the isolated-plate factors are removed, one obtains
\be
\Delta_{\rm D}(X)
=1-\left(2+\frac{4X^2}{9}\right)e^{-2X}+e^{-4X}.
\label{deltaD}
\ee
The explicit boundary matrix and its determinant are displayed in Appendix~\ref{appendix_surface}.

Two independent Dirichlet Maxwell-like modes would each contribute $1-e^{-2X}$, and hence would give $(1-e^{-2X})^2$. The difference
\be
\Delta_{\rm D}(X)-\left(1-e^{-2X}\right)^2
=-\frac{4X^2}{9}e^{-2X}
\label{deltaD_difference}
\ee
isolates the effect of the generalized profiles. The polynomial factor records the generalized normal partner generated by the repeated root: it records the coupled normal propagation of the Jordan doublet.

The determinant can also be factorized as
\be
\Delta_{\rm D}(X)=
\left[1-e^{-2(X-\theta_{\rm D})}\right]
\left[1-e^{-2(X+\theta_{\rm D})}\right],
\qquad
\theta_{\rm D}(X)=\operatorname{arcsinh}\frac{X}{3}.
\label{deltaD_factorized}
\ee
We call $\theta_{\rm D}$ the Jordan dressing function. The two quantities $X\mp\theta_{\rm D}(X)$ are effective spectral lengths: they rewrite the matrix determinant as two scalar-looking round trips. They do not change the geometric distance $h$; they summarize how the generalized profiles modify the spectrum selected by the two plates.

\subsubsection{Free boundary conditions}
When the coefficients of the physical surface action vanish,
\be
a_{\alpha,s}=b_{\alpha,s}=c_{\alpha,s}=0,
\label{vanishing_surface_couplings}
\ee
Eq.~\eqref{doublet_bc} gives the natural variational conditions
\be
2u_s'=iqv_s,
\qquad
v_s'=0
\label{free_bc}
\ee
on both plates, with the signs of the outward derivative already included in Eq.~\eqref{doublet_bc}. We call these conditions \emph{free}. The first equation retains the coupling between $u_s$ and $v_s$, while the second fixes the normal derivative of $v_s$. The finite determinant is
\be
\Delta_{\rm free}(X)
=1-\left(2+\frac{4X^2}{25}\right)e^{-2X}+e^{-4X},
\label{deltafree}
\ee
or equivalently Eq.~\eqref{deltaD_factorized} with
\be
\theta_{\rm free}(X)=\operatorname{arcsinh}\frac{X}{5}.
\label{thetafree}
\ee
For two independent massless scalar fields, ideal Dirichlet--Dirichlet and Neumann--Neumann conditions give the same Casimir coefficient. The Jordan doublet distinguishes the fixed and free variational conditions because the two components are related differently at the surface. The plate therefore probes the internal coupling of the unrestricted vector block in addition to the sign of the corresponding scalar reflection amplitude.

\subsubsection{Mixed boundary conditions}
We impose Dirichlet boundary conditions on one plate and free variational boundary conditions on the other. The finite determinant is
\be
\Delta_{\rm mix}(X)
=1+\left(\frac{34}{15}+\frac{4X^2}{15}\right)e^{-2X}+e^{-4X}.
\label{deltamixed}
\ee
The positive one-round-trip term is the matrix analogue of complementary boundary reflection. It produces a positive normalized unrestricted interaction contribution, i.e. the sign that would correspond to repulsion in a stable theory.

To compare the three cases, define
\be
I_{\mathcal B}
\equiv\int_0^\infty\dd X\,X^2\log\Delta_{\mathcal B}(X),
\qquad
\mathcal B\in\{{\rm D},{\rm free},{\rm mix}\}.
\label{doublet_integrals_definition}
\ee
Numerically,
\be
I_{\rm D}=-0.9081944\ldots,
\qquad
I_{\rm free}=-0.6718311\ldots,
\qquad
I_{\rm mix}=+0.7149576\ldots.
\label{doublet_integrals}
\ee
The Dirichlet and free coefficients of the full-theory Euclidean determinant are negative, while the mixed coefficient is positive. We use the shorthand attractive-sign and repulsive-sign for these Euclidean interaction signs, without attributing a stable-vacuum force to the full covariant theory.

\begin{figure}[H]
\centering
\includegraphics[width=0.96\textwidth]{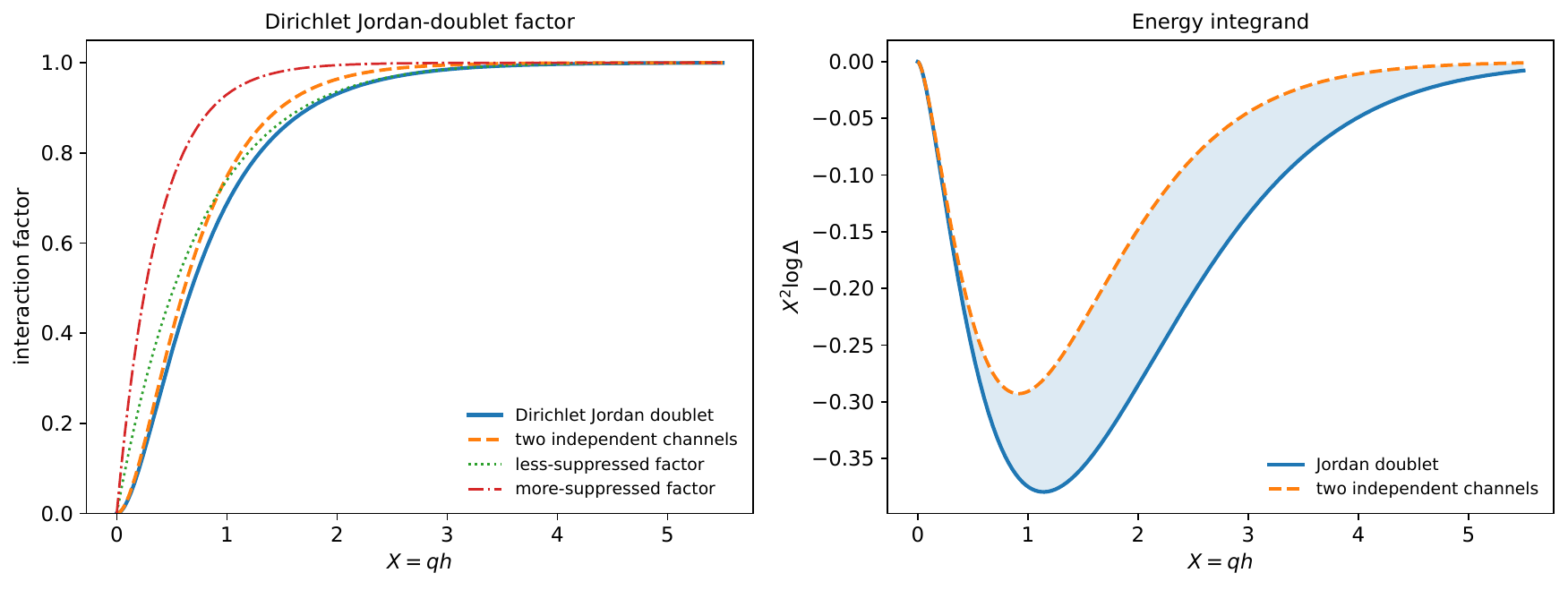}
\caption{Dirichlet response of one Jordan doublet. The left panel compares its finite interaction factor with the product of two independent Maxwell-like factors. The difference is the polynomial term in Eq.~\eqref{deltaD_difference}. The right panel shows the corresponding weighted logarithms in the interaction integral. The coupled generalized profiles make the full-theory doublet coefficient more negative than that of two independent channels.}
\label{fig:jordan_dirichlet}
\end{figure}

Figure~\ref{fig:jordan_dirichlet} compares the Dirichlet doublet with two independent Dirichlet Maxwell-like modes. The left panel displays the additional polynomial contribution in Eq.~\eqref{deltaD_difference}; the right panel shows that it makes the weighted logarithm more negative over the range of $X$ that dominates the integral. The generalized profile therefore increases the magnitude of the negative doublet interaction coefficient.

\begin{figure}[H]
\centering
\includegraphics[width=0.96\textwidth]{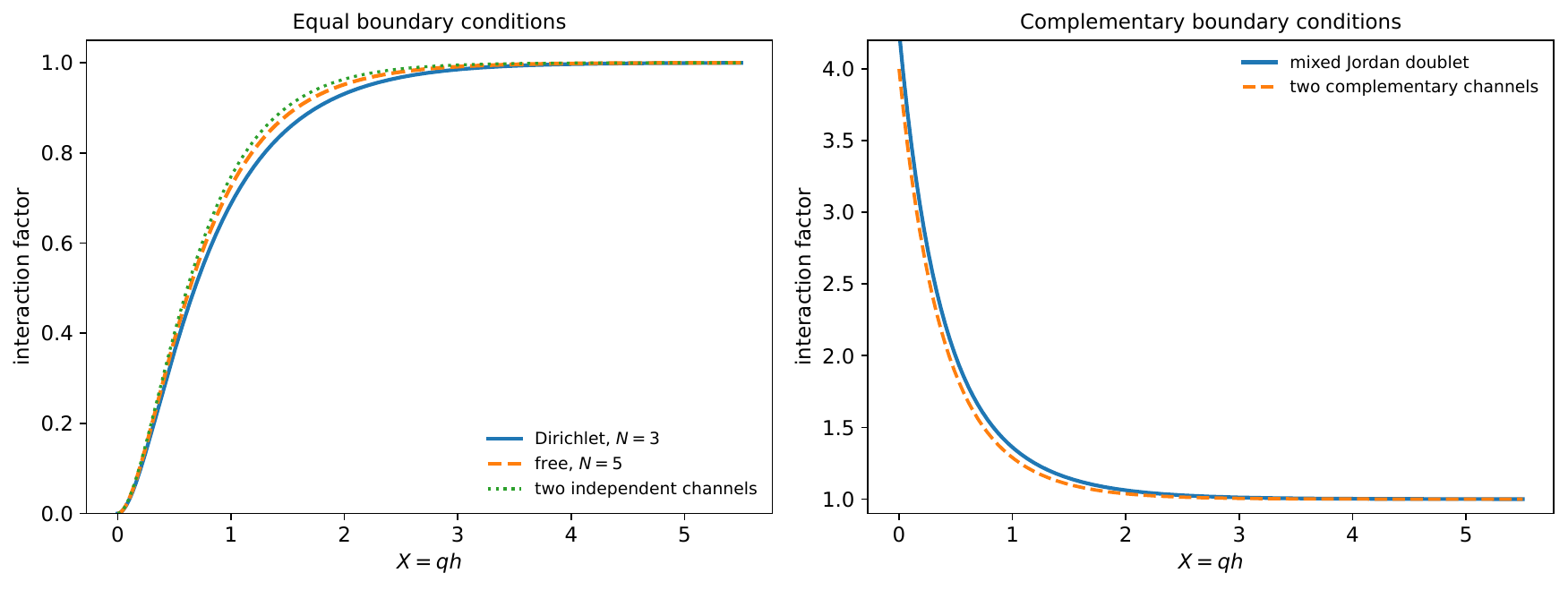}
\caption{Boundary dependence of one Jordan doublet. Dirichlet and free conditions give distinct negative coefficients, whereas two independent massless scalar channels would give the same ideal result for equal Dirichlet- or Neumann-type reflection. Complementary Dirichlet/free conditions give a positive doublet coefficient.}
\label{fig:jordan_branches}
\end{figure}

Figure~\ref{fig:jordan_branches} compares the three boundary conditions. Dirichlet and free conditions both give negative coefficients but produce distinct curves, demonstrating that the full-theory determinant depends on the relation imposed between $u_s$ and $v_s$. Mixed conditions change the sign of the normalized interaction and give a positive doublet coefficient.

\subsection{General local response}

The minimal local surface class also contains Robin and orientation-dependent response. For the diagonal entries of Eq.~\eqref{response_matrix}, locality and the restriction to at most one tangential derivative give
\be
a_{\alpha,s}(q)=\lambda_{T,\alpha}+s\gamma_{T,\alpha}q,
\qquad
c_{\alpha,s}(q)=\lambda_{N,\alpha}+s\gamma_{N,\alpha}q.
\label{ac_general}
\ee
The constants $\lambda_{T,\alpha}$ and $\lambda_{N,\alpha}$ have mass dimension one and arise from surface terms without derivatives. The dimensionless coefficients $\gamma_{T,\alpha}$ and $\gamma_{N,\alpha}$ multiply terms with one tangential derivative; in the circular basis that derivative contributes $sq$. The label $T$ refers to the tangential--tangential component $u_s=A_{m^\pm p}$, while $N$ refers to the tangential--normal component $v_s=A_{m^\pm3}$. The off-diagonal coefficient $b_{\alpha,s}$ collects local terms that mix $u_s$ and $v_s$.

For scale-free choices, all entries of $\mathcal K_{\alpha,s}$ are proportional to $q$ and introduce no new length scale. The two-plate problem can then be written in terms of a $2\times2$ round-trip matrix $\mathcal R_s(X)$. Its finite factor is
\be
\Delta_{1,s}(X)=\det\!\left[\mathbf 1_2-\mathcal R_s(X)\right].
\label{roundtrip_matrix}
\ee
For a $2\times2$ matrix, $\det(\mathbf 1_2-M)=1-\Tr M+\det M$. A complete traversal of the slab contributes $e^{-2X}$, while two traversals contribute $e^{-4X}$. The generalized profiles supply polynomial factors in $X$. The determinant therefore has the form
\be
\Delta_{1,s}(X)=1-P_s(X)e^{-2X}+Q_s e^{-4X},
\label{general_delta}
\ee
where $P_s(X)$ is a quadratic polynomial and $Q_s$ depends on the surface couplings. Appendix~\ref{appendix_surface} gives the explicit coefficients for the families used in this paper.

\section{The fractonic branch}
\label{section_fractonic_branch}

\subsection{Geometric branch condition, Gauss constraint and reduced spectrum}

Before using a $3+1$ decomposition, we specify geometrically what is meant by the fractonic branch. The full action is Lorentz covariant and contains no preferred time direction. The scalar-charge sector, by contrast, describes physics with a preferred rest frame: this is the frame in which the scalar-charge Gauss law and the associated restriction of charge mobility are defined. We encode that rest frame by a constant future-directed timelike background vector $u^\mu$ normalized by
\be
u^\mu u_\mu=-1,
\qquad
P^{\mu\nu}=\eta^{\mu\nu}+u^\mu u^\nu.
\label{branch_timelike_projector}
\ee
Here $P^{\mu\nu}$ is the spatial projector associated with $u^\mu$: it projects any spacetime vector onto the three-dimensional subspace orthogonal to the preferred time direction. The vector $u^\mu$ is part of the prescribed background geometry. It identifies the physical rest frame of the fractonic sector, in the same way that a medium defines a preferred rest frame in nonrelativistic physics. This projector has a different role from the plate-tangent projector ${\cal P}^{(\alpha)}_{\mu\nu}$ defined in Eq.~\eqref{boundary_geometry}: $P^{\mu\nu}$ separates time from space, whereas ${\cal P}^{(\alpha)}_{\mu\nu}$ separates directions tangent and normal to a plate. The full higher-rank theory is Lorentz covariant. The fractonic branch is a preferred-frame reduction, but its definition remains tensorial because $u^\mu$ transforms together with the dynamical fields under a change of coordinates.

The branch condition can be written directly in terms of the gauge-invariant field strength as
\be
u^\alpha F_{\alpha[\mu\nu]}=0,
\qquad
F_{\alpha[\mu\nu]}\equiv\frac12\left(F_{\alpha\mu\nu}-F_{\alpha\nu\mu}\right).
\label{fractonic_branch_fieldstrength}
\ee
Using Eq.~\eqref{tensor_field_strength},
\be
F_{\alpha\mu\nu}-F_{\alpha\nu\mu}
=3\left(\partial_\mu A_{\alpha\nu}-\partial_\nu A_{\alpha\mu}\right),
\label{branch_fieldstrength_identity}
\ee
so Eq.~\eqref{fractonic_branch_fieldstrength} states that the one-form $u^\alpha A_{\alpha\mu}$ is closed. Locally there is therefore a scalar field $\phi$ such that
\be
u^\alpha A_{\alpha\mu}=\partial_\mu\phi.
\label{intro_fractonic_branch}
\ee
Because $u^\mu$ is constant, this condition is preserved by the scalar gauge symmetry if
\be
\delta\phi=u^\mu\partial_\mu\Lambda.
\label{branch_phi_gauge}
\ee

For the static slab considered here, the preferred rest frame and the plate geometry are compatible. We take the plates to be at rest with respect to $u^\mu$, so that their spacelike normals are orthogonal to the timelike background,
\be
u^\mu n^{(\alpha)}_\mu=0,
\qquad \alpha\in\{0,h\}.
\label{branch_plate_orthogonality}
\ee
We may then choose a frame adapted simultaneously to the timelike background and to the plates,
\be
u^\mu=(1,0,0,0),
\qquad
n^{(0)\mu}=(0,0,0,-1),
\qquad
n^{(h)\mu}=(0,0,0,+1).
\label{branch_adapted_frame}
\ee
Only at this stage does Eq.~\eqref{intro_fractonic_branch} take the component form
\be
A_{\mu0}=\partial_\mu\phi.
\label{fractonic_branch_condition}
\ee
This is the four-dimensional scalar-charge branch used below. The corresponding preferred-frame reduction in six dimensions is analyzed in Ref.~\cite{Maggiore:2026sixdim}. In the adapted frame, Eq.~\eqref{fractonic_branch_condition} reproduces the scalar-charge component form used in the Maxwell theory of fractons \cite{Bertolini:2022ijb}. Changing coordinates transforms $u^\mu$ together with $A_{\mu\nu}$; changing the physical background $u^\mu$ instead changes the preferred rest frame that defines the reduced sector.

With spatial indices $i,j=1,2,3$, let $\epsilon_{ijk}$ be the three-dimensional Levi--Civita symbol, with $\epsilon_{123}=+1$, and define
\be
E_{ij}=F_{0ij}=\partial_0A_{ij}-\partial_i\partial_j\phi,
\qquad E_{ij}=E_{ji},
\qquad
B_{ij}=\epsilon_{ikl}\partial^k A^l{}_{j}.
\label{branch_EB}
\ee
A direct contraction gives
\be
F_{\mu\nu\rho}F^{\mu\nu\rho}\big|_{\rm FB}
=-6E_{ij}E^{ij}+6B_{ij}B^{ij},
\label{branch_Fsplit}
\ee
so Eq.~\eqref{tensor_action_lorentzian} restricted to the branch is
\be
S_{\rm FB}^{\rm L}
=g_{\rm F}\int\dd^4x\,
\left(E_{ij}E^{ij}-B_{ij}B^{ij}\right).
\label{branch_action}
\ee
Equation~\eqref{branch_action} must be supplemented by the constraints inherited from the full equations of motion. In particular, the $(0j)$ equation of the covariant theory is
\be
M_{0j}=\partial_iE^{ij}=0.
\label{branch_parent_gauss}
\ee
Substituting Eq.~\eqref{fractonic_branch_condition} into the action and then varying an unconstrained reduced functional would keep only the scalar double divergence of Eq.~\eqref{branch_parent_gauss}; it would miss its transverse projection. That autonomously varied functional is the theory analyzed in Ref.~\cite{Muniz:2026fractoncasimir}, where the transverse-vector amplitudes survive as two additional propagating modes and the spectrum splits into two propagation speeds. We therefore impose the branch on the full equations and solve this Gauss constraint before constructing the physical reduced action.

For a nonzero spatial momentum $k_i$, with $k^2\equiv k_i k_i$, introduce
\be
L_{ij}=\frac{k_i k_j}{k^2},
\qquad
T_{ij}=\delta_{ij}-L_{ij},
\label{branch_projectors}
\ee
and decompose the symmetric spatial tensor as
\be
A_{ij}
=A^{\rm TT}_{ij}+ik_iV^T_j+ik_jV^T_i
+\frac12T_{ij}\sigma+L_{ij}\rho,
\label{branch_decomposition}
\ee
with $k^iA^{\rm TT}_{ij}=0$, $\delta^{ij}A^{\rm TT}_{ij}=0$ and $k^iV^T_i=0$. A dot denotes a derivative with respect to Lorentzian time. The electric tensor is
\be
E_{ij}
=\dot A^{\rm TT}_{ij}+ik_i\dot V^T_j+ik_j\dot V^T_i
+\frac12T_{ij}\dot\sigma
+L_{ij}\left(\dot\rho+k^2\phi\right).
\label{branch_E_decomposition}
\ee
The decomposition in Eq.~\eqref{branch_decomposition} is simply a bookkeeping device for the six independent components of the spatial symmetric tensor. It contains two TT amplitudes, two transverse-vector amplitudes, one transverse trace $\sigma$ and one longitudinal scalar $\rho$:
\be
6=2_{\rm TT}+2_{V^T}+1_{\sigma}+1_{\rho}.
\label{branch_component_count_before_gauss}
\ee
No physical conclusion has yet been drawn from this decomposition alone. The reduction occurs when the Gauss equation \eqref{branch_parent_gauss} is imposed. It separates into
\be
\dot V^T_i=0,
\qquad
\dot\rho+k^2\phi=0.
\label{branch_constraints}
\ee
The first equation removes the transverse-vector amplitudes from the radiative spectrum. In the source-free theory considered here the statement is in fact stronger than the absence of propagation. The remaining equations of the full theory in the same sector are
\be
2\ddot V^T_i+k^2V^T_i=0,
\label{branch_VT_second_order}
\ee
and the Gauss constraint $\dot V^T_i=0$ gives $\ddot V^T_i=0$, so that
\be
k^2V^T_i=0,
\qquad
V^T_i=0
\quad\text{for }k\neq0.
\label{branch_VT_vanishes}
\ee
At nonzero spatial momentum the two equations together therefore enforce $V^T_i=0$ identically. No special choice of initial data is required, and time evolution cannot regenerate the transverse-vector sector. The fractonic branch is in this sense a dynamically consistent restriction of the free covariant theory. The statement concerns the source-free quadratic theory; generic sources or interactions need not preserve the same truncation.

The second equation fixes the longitudinal scalar in terms of $\phi$ and likewise does not introduce an additional propagating mode; the remaining equation $M_{33}=0$ is its time derivative and is not independent. It is on the branch that $E_{ij}$ of Eq.~\eqref{branch_EB} becomes the electric tensor of scalar-charge fracton electrodynamics. Taking a further divergence of the inherited vector constraint \eqref{branch_parent_gauss} gives the source-free scalar-charge Gauss law. In the autonomous scalar-charge theory that double-divergence equation is the fundamental constraint, and with a scalar charge density it becomes $\partial_i\partial_jE^{ij}=\rho$, which is what enforces the conservation of charge and dipole moment \cite{Pretko:2020cko,Pretko:2016kxt}. The radiative variables are therefore
\be
A^{\rm TT}_{ij},
\qquad
\sigma,
\label{branch_physical_fields}
\ee
with
\be
N_{\rm FB}=2_{\rm TT}+1_{\rm tr}=3.
\label{branch_dof_count}
\ee

For a momentum chosen along the third spatial direction, the physical field is a transverse $2\times2$ symmetric tensor. Its norm splits as
\be
A_{AB}A^{AB}=A^{\rm TT}_{AB}A_{\rm TT}^{AB}+\frac12\sigma^2,
\qquad A,B=1,2.
\label{branch_norm_split}
\ee
After constant field rescalings, $H^{\rm TT}_{ij}=\sqrt{2g_{\rm F}}\,A^{\rm TT}_{ij}$ and $\tau=\sqrt{g_{\rm F}}\,\sigma$, the reduced action is
\be
S_{\rm FB,red}^{\rm L}
=\frac12\int\dd^4x\,
\left[
\dot H^{\rm TT}_{ij}\dot H_{\rm TT}^{ij}
-\partial_kH^{\rm TT}_{ij}\partial^kH_{\rm TT}^{ij}
+\dot\tau^2-\partial_k\tau\partial^k\tau
\right].
\label{branch_reduced_action}
\ee
The canonical momenta of the reduced fields are $\Pi^{\rm TT}_{ij}=\dot H^{\rm TT}_{ij}$ and $\pi_\tau=\dot\tau$. The corresponding Hamiltonian is
\be
H_{\rm FB,red}
=\frac12\int\dd^3x\,
\left[
\Pi^{\rm TT}_{ij}\Pi_{\rm TT}^{ij}
+\partial_kH^{\rm TT}_{ij}\partial^kH_{\rm TT}^{ij}
+\pi_\tau^2+\partial_k\tau\partial^k\tau
\right]\ge0.
\label{branch_positive_hamiltonian}
\ee
Thus the branch removes precisely the indefinite transverse-vector sector of Section~\ref{section_parent_vector_stability} and retains three positive-energy massless modes.

\subsection{Normal propagation in the slab}

The branch selects physical time through the background $u^\mu$, so its slab decomposition is performed directly in the preferred rest frame rather than inferred by deleting rows from the Euclidean helicity decomposition of the full theory. Instead we use precisely the frame in Eq.~\eqref{branch_adapted_frame}, adapted simultaneously to the preferred rest frame and to the plate normals. This keeps the two geometric structures conceptually separate even though they are represented by simple coordinate axes. After Wick rotation, rotate the spatial momentum parallel to the plates to $k_2=0$ and write
\be
q^2=\zeta^2+k^2,
\qquad k\equiv k_1.
\label{branch_q}
\ee
For $\zeta\neq0$ a branch-preserving gauge transformation sets $\phi=0$. The Gauss equation of the full theory then imposes spatial transversality,
\be
ikA_{1j}+\partial_3A_{3j}=0.
\label{branch_transversality_slab}
\ee
On this constrained subspace the remaining spatial equations reduce to
\be
\left(-\partial_3^2+q^2\right)A_{ij}=0.
\label{branch_helmholtz_tensor}
\ee
Three local components can be chosen as independent normal amplitudes,
\be
f=A_{11},
\qquad
g=A_{12},
\qquad
r=A_{22}.
\label{branch_independent_amplitudes}
\ee
Each satisfies Eq.~\eqref{intro_helmholtz}. The remaining components are fixed by Gauss; for example,
\be
A_{13}=-\frac{ik}{q^2}f',
\qquad
A_{23}=-\frac{ik}{q^2}g',
\qquad
A_{33}=-\frac{k^2}{q^2}f.
\label{branch_dependent_components}
\ee
There are therefore three ordinary growing/decaying pairs and no generalized normal partners. An independent symbolic rank calculation of the full equations restricted to the branch gives one pure-gauge direction off shell and three physical null directions on $\Box=0$, in agreement with Eqs.~\eqref{branch_dof_count}--\eqref{branch_helmholtz_tensor}. The subset at zero Euclidean frequency is obtained by continuity of the frequency integral; constrained static modes do not add an independent radiative determinant.

\subsection{Branch-compatible ideal reflectors}

The branch is a restriction of the solution space of the full theory, and the natural boundary conditions of the unrestricted tensor problem do not in general remain admissible after that restriction. The reason can be seen directly. For the amplitudes in Eqs.~\eqref{branch_independent_amplitudes} and \eqref{branch_dependent_components}, the normal momenta of the full theory of Eq.~\eqref{boundary_momenta} contain, at either plate,
\be
\pi_{11}=\frac{\zeta^2}{q^2}f',
\qquad
\pi_1=-\frac{ik\zeta^2}{q^2}f,
\label{branch_parent_bc_f}
\ee
and
\be
\pi_{12}=\left(1-\frac{k^2}{2q^2}\right)g',
\qquad
\pi_2=-ikg,
\qquad
\pi_{22}=r'.
\label{branch_parent_bc_gr}
\ee
Hence, for generic $\zeta k\neq0$, the simultaneous unrestricted conditions $\pi_{11}=\pi_1=0$ force both $f'=0$ and $f=0$ at the same plate; similarly, $\pi_{12}=\pi_2=0$ force $g'=0$ and $g=0$. This overconstrains the reduced problem. Each reduced amplitude satisfies a second-order Helmholtz equation, so a separated self-adjoint slab problem supplies one independent linear boundary condition per amplitude at each plate. Requiring both the value and the normal derivative of the same amplitude at one plate imposes two conditions where only one is available and generically removes the solution rather than selecting a reflection law. The extra conditions arise because the unrestricted variation treats tensor directions as independent that are tied together after the branch and Gauss projection. The physical branch therefore requires its own self-adjoint boundary domain after the full-theory constraints have been solved.

Let
\be
\boldsymbol\varphi=(\varphi_1,\varphi_2,\varphi_3)^T
\label{branch_canonical_vector}
\ee
be any orthonormal set of the two TT amplitudes and the canonical trace scalar $\tau$. Each component obeys Eq.~\eqref{intro_helmholtz}. A general reduced reflection problem is described by $3\times3$ reflection matrices $R_0$ and $R_h$ associated with the lower and upper plates, respectively. With $\mathbb I_3$ denoting the $3\times3$ identity matrix, the finite factor is
\be
\Delta_{\rm FB}(q,h)
=\det\!\left[\mathbb I_3-R_0R_h e^{-2qh}\right].
\label{branch_general_determinant}
\ee
We use a superscript $(+)$ for equal ideal reflection and $(-)$ for complementary ideal reflection. For equal Dirichlet--Dirichlet or Neumann--Neumann reflection, $R_0R_h=+\mathbb I_3$, so
\be
\Delta_{\rm FB}^{(+)}(X)=\left(1-e^{-2X}\right)^3.
\label{branch_delta_plus}
\ee
For complementary ideal Dirichlet--Neumann reflection, $R_0R_h=-\mathbb I_3$, and
\be
\Delta_{\rm FB}^{(-)}(X)=\left(1+e^{-2X}\right)^3.
\label{branch_delta_minus}
\ee
The fixed-value realization has a simple local expression in the frame adapted to the time direction and the plates: fixing the three spatially tangential components $A_{AB}$, $A,B=1,2$, fixes the three independent branch amplitudes. By contrast, after Gauss projection a canonical Neumann condition on $\boldsymbol\varphi$ is generally nonlocal when rewritten in terms of the unreduced spacetime tensor. We therefore call Eqs.~\eqref{branch_delta_plus} and \eqref{branch_delta_minus} \emph{reduced ideal reflectors}. The reduced Neumann condition is a boundary condition on the three physical branch modes; it is distinct from the vanishing-coupling natural condition of the full tensor theory.

Using the integrals $I_+$ and $I_-$ defined in Eqs.~\eqref{Iplus} and \eqref{Iminus}, the physical branch energies are
\be
\frac{E_{\rm FB}^{(+)}(h)}{\Acal}
=\frac{3I_+}{4\pi^2h^3}
=-\frac{\pi^2}{480h^3},
\label{branch_equal_energy}
\ee
and
\be
\frac{E_{\rm FB}^{(-)}(h)}{\Acal}
=\frac{3I_-}{4\pi^2h^3}
=\frac{7\pi^2}{3840h^3}.
\label{branch_complementary_energy}
\ee
Because Maxwell theory contains two physical channels with the same ideal reflection products,
\be
\frac{E_{\rm FB}^{(\pm)}}{E_{\rm EM}^{(\pm)}}=\frac32.
\label{branch_maxwell_ratio}
\ee
Differentiating with respect to $h$ gives the corresponding physical forces,
\be
\frac{F_{\rm FB}^{(+)}(h)}{\Acal}=-\frac{\pi^2}{160h^4},
\qquad
\frac{F_{\rm FB}^{(-)}(h)}{\Acal}=+\frac{7\pi^2}{1280h^4}.
\label{branch_forces}
\ee
The branch result is therefore a pure physical-mode multiplicity for ideal reflectors. The boundary-dependent Jordan enhancement belongs to the full covariant theory and disappears together with its unstable vector sector.

We do not extend the branch calculation to a family of non-ideal reflection matrices at this stage. Reflection coefficients that are not derived from a surface action compatible with the reduced phase space would introduce free phenomenological parameters, and the preceding paragraphs have shown that the natural conditions of the full tensor theory cannot simply be projected onto the branch: they overconstrain it. The universally controlled branch statement available here is therefore the ideal-reflector one, and the construction of local, branch-compatible and possibly dispersive surface response is treated as the separate problem it is.

\section{Casimir comparison: Maxwell, full covariant theory and fractonic branch}
\label{section_results}

The common normalization is now fixed. The radial reduction and the two ideal-reflection integrals $I_+$ and $I_-$ were derived in Section~\ref{section_maxwell}; $I_+$ corresponds to equal ideal reflection and $I_-$ to complementary ideal reflection. We can therefore compare the three theories without introducing any new kinematic quantity.

\subsection{Euclidean determinant of the unrestricted covariant theory}

After the BRS cancellation of the gauge-dependent sector, the full higher-rank theory contains three independent Maxwell-like modes and two degenerate two-component Jordan doublets. Denote its normalized Euclidean interaction functional by $\mathcal E_{\rm U}$, where the subscript U stands for \emph{unrestricted}. For a boundary class $\mathcal B$,
\be
\frac{\mathcal E_{\rm U}^{(\mathcal B)}(h)}{\Acal}
=\frac12\int\frac{\dd^3p}{(2\pi)^3}
\left[
\sum_{j=1}^{3}\log\Delta_{{\rm M},j}^{(\mathcal B)}(X)
+\sum_{s=\pm1}\log\Delta_{1,s}^{(\mathcal B)}(X)
\right].
\label{parent_complete_energy}
\ee
Here $\Delta_{{\rm M},j}$ is the one-channel factor of an independent Maxwell-like mode, while $\Delta_{1,s}$ is the two-component Jordan factor defined in Section~\ref{section_boundary_doublets}. The notation $\mathcal E_{\rm U}$ is deliberate: Section~\ref{section_parent_vector_stability} showed that the transverse-vector sector has no Hamiltonian bounded below. Equation~\eqref{parent_complete_energy} is therefore a normalized Euclidean spectral functional, not the ground-state energy of a stable Lorentzian theory.

For the three representative local boundary problems studied in Section~\ref{section_boundary_doublets},
\be
\begin{aligned}
\frac{\mathcal E_{\rm U}^{({\rm D})}}{\Acal}
&=\frac{1}{4\pi^2h^3}\left(3I_++2I_{\rm D}\right),\\
\frac{\mathcal E_{\rm U}^{({\rm free})}}{\Acal}
&=\frac{1}{4\pi^2h^3}\left(3I_++2I_{\rm free}\right),\\
\frac{\mathcal E_{\rm U}^{({\rm mix})}}{\Acal}
&=\frac{1}{4\pi^2h^3}\left(3I_-+2I_{\rm mix}\right).
\end{aligned}
\label{parent_three_energies}
\ee
Using the Jordan integrals defined in Eq.~\eqref{doublet_integrals}, define the dimensionless ratio $\mathcal R_{\rm U,\mathcal B}\equiv\mathcal E_{\rm U}^{(\mathcal B)}/E_{\rm EM}^{(\mathcal B)}$ for the Maxwell configuration with the same ideal reflection product. The three values are
\be
\begin{aligned}
\mathcal R_{\rm U,D}&=4.85646\ldots,\\
\mathcal R_{\rm U,free}&=3.98292\ldots,\\
\mathcal R_{\rm U,mix}&=4.51978\ldots.
\end{aligned}
\label{parent_ratios}
\ee
These numbers characterize the Euclidean boundary problem of the unrestricted theory. They are not physical Casimir-energy ratios, because the Lorentzian vector sector is unstable. Their value is diagnostic: if the seven gauge-invariant components behaved as seven independent scalar channels, the corresponding Maxwell ratio would be $7/2$; the different values in Eq.~\eqref{parent_ratios} measure the effect of the coupled Jordan propagation and reflection.

\subsection{Physical fractonic branch}

The scalar-charge branch has a different status. The Gauss constraint removes the two transverse-vector pairs, and the reduced Hamiltonian contains the three positive-energy modes derived in Section~\ref{section_fractonic_branch}. For ideal equal or complementary reflectors, each of those modes has the same reflection product as a Maxwell polarization. Hence
\be
\frac{E_{\rm FB}^{(+)}}{E_{\rm EM}^{(+)}}
=\frac{E_{\rm FB}^{(-)}}{E_{\rm EM}^{(-)}}
=\frac32.
\label{physical_branch_ratios}
\ee
The corresponding energies, already obtained from the reduced determinant in Section~\ref{section_fractonic_branch}, are
\be
\frac{E_{\rm FB}^{(+)}}{\Acal}=-\frac{\pi^2}{480h^3},
\qquad
\frac{E_{\rm FB}^{(-)}}{\Acal}=\frac{7\pi^2}{3840h^3}.
\label{physical_branch_energies}
\ee
The factor $3/2$ is therefore not an assumption based on the number of tensor components. It is the result of the dynamical restriction: the full equations first remove the unstable vector sector, after which the reduced action contains exactly three stable massless channels.

\begin{table}[H]
\centering
\small
\begin{tabular}{@{}>{\raggedright\arraybackslash}p{0.20\textwidth}>{\raggedright\arraybackslash}p{0.29\textwidth}>{\raggedright\arraybackslash}p{0.22\textwidth}>{\raggedright\arraybackslash}p{0.19\textwidth}@{}}
\toprule
system & physical/spectral content & representative result & interpretation \\
\midrule
Maxwell & two stable photon modes & $E_{\rm EM}^{(+)}/\Acal=-\pi^2/(720h^3)$ & physical vacuum energy \\
full covariant theory & three Maxwell-like modes plus two two-component Jordan doublets & $\mathcal R_{\rm U,D}=4.856$, $\mathcal R_{\rm U,free}=3.983$ & Euclidean spectral functional; vector Hamiltonian indefinite \\
fractonic branch & preferred-frame $2_{\rm TT}+1_{\rm tr}$ sector, positive Hamiltonian & $E_{\rm FB}^{(+)}/E_{\rm EM}^{(+)}=3/2$ & physical fracton Casimir energy \\
\bottomrule
\end{tabular}
\caption{The three systems compared in this work. The full covariant theory contains a transverse-vector sector with Jordan dynamics and indefinite Lorentzian energy. The scalar-charge restriction removes that sector, leaving three stable massless modes whose ideal Casimir energy is three halves of the Maxwell value.}
\label{table:three_theories}
\end{table}

The comparison has a direct physical interpretation. Maxwell theory tests two stable photon modes. The unrestricted higher-rank determinant also sees the coupled vector sector, but that sector is not part of a stable Lorentzian vacuum. The fractonic branch keeps only the three stable modes compatible with the scalar-charge constraints. In this sense the Casimir calculation makes visible, at the level of vacuum propagation between plates, the same restriction of physical motion that distinguishes the fractonic sector from the full covariant theory.

\section{Orientation-dependent response of the full covariant theory}
\label{section_odd}

\subsection{Why the orientation of the plates matters}

The orientation tensor on a plate is defined by Eq.~\eqref{boundary_geometry}. When the same tangential coordinates are used on both plates, the outward normals point in opposite directions and the induced orientation tensors have opposite signs. A local surface interaction containing $\varepsilon^{(\alpha)}_{\mu\nu\rho}$ therefore appears with opposite coefficients on the two plates even when the plates have the same physical properties. We refer to this configuration as the \emph{orientation-odd flip configuration}. A related orientation-based construction appears in the Maxwell--Chern--Simons strip \cite{Maggiore:2026agw}; higher-rank Chern--Simons response also provides a natural context for orientation-sensitive tensor couplings \cite{Bertolini:2024hall}. Throughout this section, stability refers only to the absence of negative isolated-surface modes and slab zeros of the Euclidean boundary determinant. It does not remove the Lorentzian bulk vector instability established in Section~\ref{section_parent_vector_stability}.

Consider first one independent Maxwell-like mode in a circular polarization basis, with $s=\pm1$ and a dimensionless orientation-odd coupling $\gamma$. The two plates have
\be
K_{0,s}=s\gamma q,
\qquad
K_{h,s}=-s\gamma q.
\label{scalar_flip_K}
\ee
Using Eq.~\eqref{maxwell_reflection},
\be
r_{0,s}=\frac{1-s\gamma}{1+s\gamma},
\qquad
r_{h,s}=\frac{1+s\gamma}{1-s\gamma},
\qquad
r_{0,s}r_{h,s}=1.
\label{scalar_flip_reflections}
\ee
The round-trip factor remains $1-e^{-2X}$. This cancellation applies to each independently propagating mode. It therefore leaves unchanged both photon polarizations and the three independent Maxwell-like modes of the full theory.

By contrast, a Jordan doublet has a matrix-valued reflection amplitude. Reversing the plate orientation changes the corresponding matrix round trip, so the doublet retains a dependence on the orientation-odd surface couplings. Plate orientation therefore leaves the independent Maxwell and Maxwell-like modes unchanged while modifying the coupled vector contribution of the full theory.

\subsection{Equal diagonal orientation-odd couplings}

Everything in this section concerns the Euclidean boundary spectral problem of the full covariant theory. It is a statement about admissible boundary domains for that determinant, and it is logically independent of the Lorentzian bulk instability established in Section~\ref{section_parent_vector_stability}; neither implies the other. In particular, no result of this section is a claim about a physical force. 

We first set the off-diagonal response to zero and choose equal orientation-odd couplings for the two diagonal entries of Eq.~\eqref{response_matrix}. For a fixed doublet,
\be
b_{0,s}=b_{h,s}=0,
\qquad
a_{0,s}=c_{0,s}=s\gamma q,
\qquad
a_{h,s}=c_{h,s}=-s\gamma q.
\label{gamma_surface_coefficients}
\ee
The parameter $\gamma$ is dimensionless. It multiplies the diagonal terms $|u_s|^2$ and $|v_s|^2$ in the surface action and changes sign when the outward normal is reversed. Equation~\eqref{doublet_bc} becomes
\be
\begin{aligned}
-(2u_s'-iqv_s)+s\gamma q\,u_s&=0,
&-v_s'+s\gamma q\,v_s&=0
&&\text{at }\Sigma_0,\\
 +(2u_s'-iqv_s)-s\gamma q\,u_s&=0,
& v_s'-s\gamma q\,v_s&=0
&&\text{at }\Sigma_h.
\end{aligned}
\label{gamma_boundary_conditions}
\ee
Substituting the four profiles in Eq.~\eqref{doublet_basis} into these four equations gives $Q=1$ in Eq.~\eqref{general_delta} and
\be
P_\gamma(X)
=2+4X^2\frac{1-\gamma^2}{25-9\gamma^2}.
\label{Pgamma}
\ee
The finite determinant is therefore
\be
\Delta_\gamma(X)
=1-\left[2+4X^2\frac{1-\gamma^2}{25-9\gamma^2}\right]e^{-2X}+e^{-4X}.
\label{delta_gamma_explicit}
\ee
It is convenient to define the dimensionless parameter
\be
N_{\rm eff}^2(\gamma)=\frac{25-9\gamma^2}{1-\gamma^2},
\label{Neff}
\ee
so that
\be
\Delta_\gamma(X)
=1-\left(2+\frac{4X^2}{N_{\rm eff}^2(\gamma)}\right)e^{-2X}+e^{-4X}.
\label{delta_gamma}
\ee
At $\gamma=0$, $N_{\rm eff}=5$ and Eq.~\eqref{delta_gamma} reduces to the free determinant \eqref{deltafree}. As $|\gamma|\to1$, $N_{\rm eff}\to\infty$ and the polynomial term tends to zero,
\be
\Delta_\gamma(X)\longrightarrow\left(1-e^{-2X}\right)^2.
\label{gamma_scalar_limit}
\ee
The orientation-dependent surface response therefore suppresses the part of the Casimir determinant that detects the generalized Jordan profiles. The bulk doublet remains coupled; it is the boundary sensitivity to that coupling that becomes weaker.

\subsection{Boundary spectrum of the isolated-plate reference configuration}

The normalized unrestricted interaction functional is defined by subtracting the limit of infinitely separated plates, Eq.~\eqref{subtraction}. Its reference configuration therefore contains two isolated plates. Before analytically continuing the logarithm of the two-plate determinant, one must require the isolated-surface Euclidean operator to have no negative localized eigenvalue. This is a boundary spectral condition; it is logically separate from the Lorentzian bulk instability of the full covariant theory.

Consider the half-space $x^3\geq0$ and the eigenvalue problem
\be
\mathcal M_1(\partial_3)
\begin{pmatrix}u\\v\end{pmatrix}
=\lambda
\begin{pmatrix}u\\v\end{pmatrix},
\label{surface_eigenvalue_problem}
\ee
with the boundary conditions generated by Eq.~\eqref{gamma_surface_coefficients} and with a profile that decays as $x^3\to\infty$. For $\gamma>0$, the solution
\be
\begin{pmatrix}u\\v\end{pmatrix}
=\begin{pmatrix}-i\\\gamma\end{pmatrix}e^{-\gamma qx^3}
\label{surface_mode}
\ee
satisfies the boundary conditions and has eigenvalue
\be
\lambda_{\rm surf}=q^2(1-\gamma^2).
\label{surface_eigenvalue}
\ee
Changing the sign of $\gamma$ exchanges the plate and helicity sector in which the localized solution occurs; the boundary-spectral condition depends only on $|\gamma|$. The strictly boundary-stable interval is
\be
|\gamma|<1.
\label{gamma_stable}
\ee
Within this interval the orientation-dependent response continuously reduces the magnitude of the negative unrestricted interaction functional while preserving its sign. To describe the analytic continuation beyond the boundary-stable interval, define the doublet integral
\be
I_\gamma(\gamma)
\equiv\int_0^\infty\dd X\,X^2\log\Delta_\gamma(X).
\label{Igamma_definition}
\ee
The continued doublet contribution changes sign at $\gamma^*$, while the complete unrestricted interaction functional changes sign at $\gamma^{**}$:
\be
I_\gamma(\gamma^*)=0,
\qquad
3I_++2I_\gamma(\gamma^{**})=0,
\label{gamma_crossing_definitions}
\ee
with
\be
\gamma^*\simeq1.4412,
\qquad
\gamma^{**}\simeq1.5173.
\label{formal_crossings}
\ee
Both values occur after the surface eigenvalue has become negative. They belong to the analytic continuation of a boundary-spectrally unstable problem and are not part of the admissible boundary domain of the full theory.

\begin{figure}[H]
\centering
\includegraphics[width=0.96\textwidth]{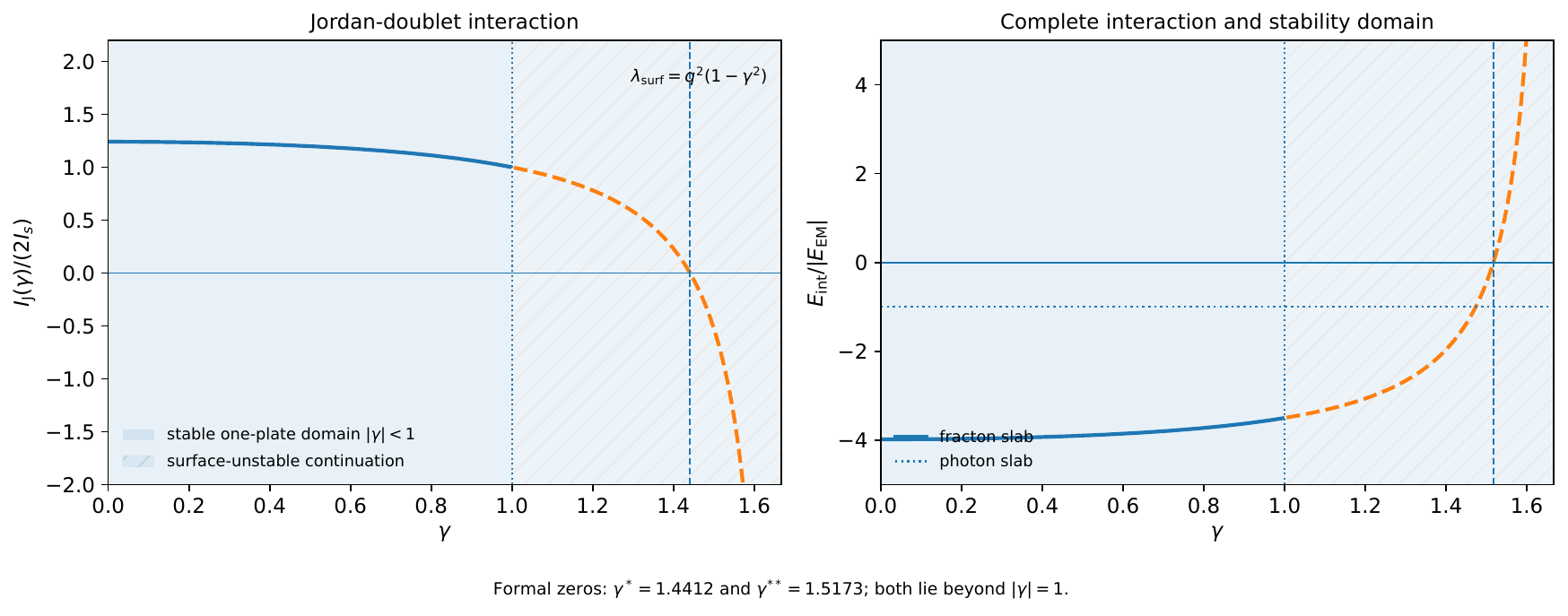}
\caption{Orientation-dependent diagonal response. The Maxwell interaction and the three independent Maxwell-like modes do not depend on $\gamma$ because the scalar reflection amplitudes cancel in the round trip. The Jordan doublets retain a $\gamma$ dependence. The unhatched interval $|\gamma|<1$ is the boundary-stable isolated-plate domain. The formal zeros of the analytically continued unrestricted interaction functional lie outside this interval, after a negative Euclidean surface mode has appeared.}
\label{fig:flip}
\end{figure}

Figure~\ref{fig:flip} displays both parts of this conclusion. In the interval $|\gamma|<1$, the Maxwell energy and the three independent Maxwell-like modes remain fixed, while the doublet contribution changes continuously. The formal sign changes lie in the shaded region where the isolated plate already supports a negative Euclidean mode.

\subsection{Orientation-odd mixing of the two doublet components}

A second family acts through the off-diagonal entry of the surface response matrix. We set the diagonal coefficients to zero and, for one doublet, choose
\be
 a_{0,s}=c_{0,s}=a_{h,s}=c_{h,s}=0,
\qquad
 b_{0,s}=i\beta q,
\qquad
 b_{h,s}=-i\beta q.
\label{beta_surface_coefficients}
\ee
The opposite-helicity doublet has the complex-conjugate coefficients and the same determinant. The dimensionless parameter $\beta$ therefore measures a local orientation-odd mixing between $u_s$ and $v_s$ at the plate. For the displayed doublet, the boundary conditions are
\be
\begin{aligned}
-(2u'-iqv)+i\beta q\,v&=0,
&-v'-i\beta q\,u&=0
&&\text{at }\Sigma_0,\\
 +(2u'-iqv)-i\beta q\,v&=0,
& v'+i\beta q\,u&=0
&&\text{at }\Sigma_h.
\end{aligned}
\label{beta_boundary_conditions}
\ee
Substitution of the normal basis into these equations gives
\be
\Delta_\beta(X)
=1-\left(2+\frac{4X^2}{N_\beta^2}\right)e^{-2X}+e^{-4X},
\label{delta_beta}
\ee
with
\be
N_\beta^2=\frac{(3\beta+5)^2}{(\beta-1)^2}.
\label{Nbeta}
\ee
The family passes through several useful values:
\be
\beta=-1:\ N_\beta=1,
\qquad
\beta=-\frac13:\ N_\beta=3,
\qquad
\beta=0:\ N_\beta=5,
\qquad
\beta\to1^-:\ N_\beta\to\infty.
\label{Nbeta_values}
\ee
Thus a continuous boundary mixing interpolates between different strengths of the polynomial Jordan contribution, including the Dirichlet and free values.

The boundary-spectral analysis has two parts. First, the half-space problem tests whether an isolated plate creates a negative localized Euclidean eigenmode. The determinant of the decaying half-space solutions gives
\be
-\frac53<\beta<1.
\label{beta_oneplate}
\ee
Second, the two-plate factor $\Delta_\beta(X)$ must remain positive for every $X>0$ so that the Euclidean logarithm is real and the slab fluctuation problem has no zero inside the integration range. This requirement gives
\be
-1\leq\beta<1.
\label{beta_slab}
\ee
The boundary-stable domain is the intersection of the two conditions, namely Eq.~\eqref{beta_slab}. The derivation is given in Appendix~\ref{appendix_stability}.

Figure~\ref{fig:beta} shows how the boundary mixing changes the doublet determinant within the boundary-stable domain. The left panel displays the interpolation of $N_\beta$; the right panel shows the corresponding negative contribution to the normalized full-theory interaction functional. The comparison again has no Maxwell analogue: the independently propagating photon channels remain unchanged, whereas the Jordan doublets of the full theory retain information about the matrix structure of the surface response.

\begin{figure}[H]
\centering
\includegraphics[width=0.96\textwidth]{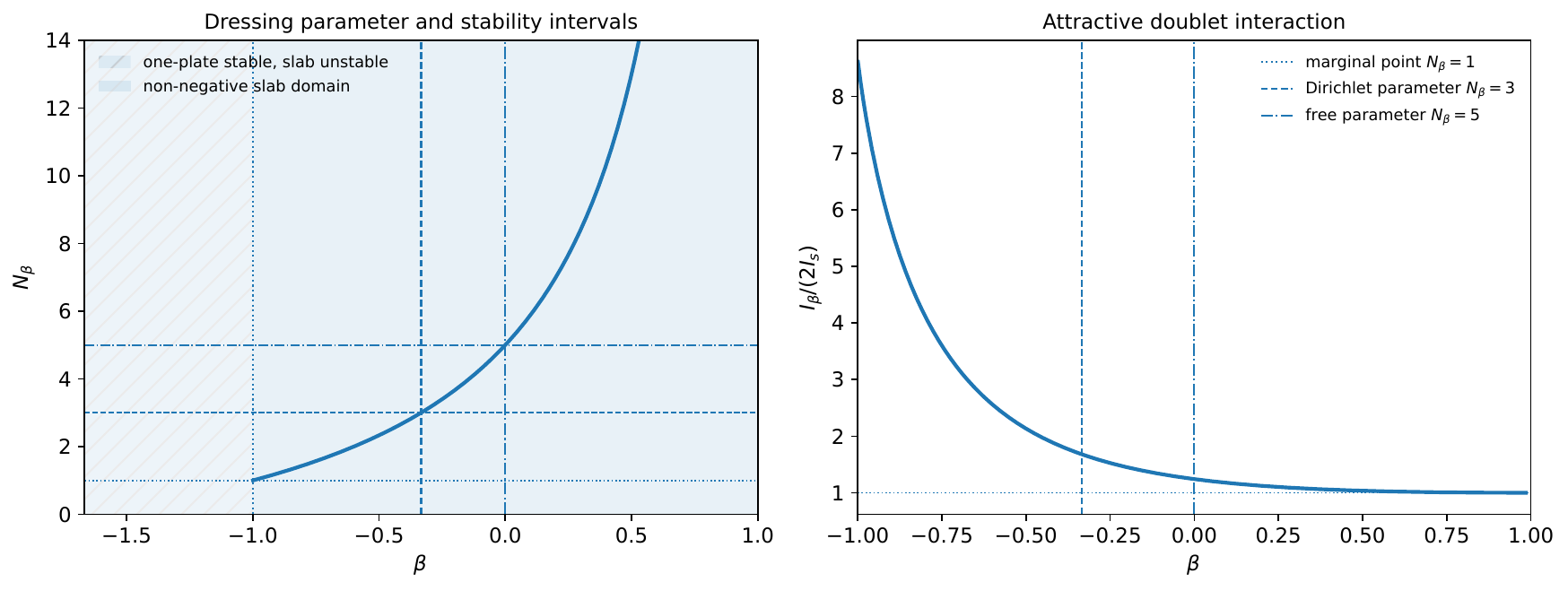}
\caption{Orientation-odd off-diagonal mixing. The half-space and two-plate positivity conditions select the boundary-stable interval $-1\leq\beta<1$. Within it, the parameter $N_\beta$ continuously changes the polynomial Jordan contribution and passes through the Dirichlet value $N_\beta=3$ and the free value $N_\beta=5$. The normalized unrestricted interaction functional remains negative throughout this interval.}
\label{fig:beta}
\end{figure}

The half-space and slab tests answer two Euclidean boundary questions. The half-space calculation excludes a negative isolated-surface eigenvalue, while the slab determinant verifies that the two-plate boundary problem remains regular for every tangential momentum. Their intersection is therefore the domain in which the normalized unrestricted determinant is real and free of boundary-induced zeros. It does not turn the full Lorentzian theory into a stable vacuum theory.

\section{Conclusions and outlook}
\label{section_conclusions}

The purpose of this work was to separate three questions that can easily be confused: the standard Casimir problem of Maxwell electrodynamics, the Euclidean boundary problem of the full Lorentz-covariant higher-rank theory, and the physical Casimir effect of its scalar-charge fractonic branch. Treating them in the same slab geometry makes clear which features come from the higher-rank covariant completion and which survive after the restriction that defines the fractonic sector.

The unrestricted higher-rank theory has a boundary spectrum that is substantially richer than Maxwell theory. In addition to three independent Maxwell-like modes, it contains two coupled vector pairs with defective repeated roots. Their generalized normal profiles produce polynomial contributions to the finite two-plate determinant, and local surface actions can probe this matrix structure through fixed, free, mixed and orientation-dependent boundary conditions. These Euclidean determinants are well-defined spectral quantities. The Lorentzian analysis, however, shows that the same vector sector has an indefinite Hamiltonian. The large numerical coefficients found for the unrestricted determinant therefore characterize the covariant boundary problem; they are not interpreted as the Casimir energy of a stable vacuum.

The scalar-charge branch removes precisely that vector sector. Its definition requires a preferred timelike background, which identifies the physical rest frame of the fractonic theory while allowing the reduction itself to be written tensorially \cite{Maggiore:2026sixdim}. Keeping the Gauss constraint of the full theory together with the branch condition eliminates the transverse-vector amplitudes and fixes the longitudinal scalar. The remaining radiative content consists of two transverse-traceless tensor modes and one scalar trace mode, all massless and with positive energy.

This reduction also provides the direct link with the defining physics of fractons. In the scalar-charge theory, the double-divergence Gauss law implies conservation of both charge and dipole moment; an isolated charge cannot move without changing the dipole moment. In the source-free problem studied here, the same constraint structure restricts the radiative sector before the Casimir determinant is formed. The vacuum interaction is therefore sensitive to a consequence of the same kinematics that produces restricted mobility when charges are present: fewer physical sectors remain available to propagate between the plates.

For ideal reflectors, the consequence is exact. The three stable modes of the fractonic branch propagate independently with the same normal equation as a Maxwell polarization. The branch Casimir energy is therefore three halves of the corresponding Maxwell value, for both equal and complementary ideal boundary conditions. In particular,
\be
\frac{E_{\rm FB}^{(+)}}{\Acal}=-\frac{\pi^2}{480h^3},
\qquad
\frac{E_{\rm FB}^{(-)}}{\Acal}=\frac{7\pi^2}{3840h^3},
\qquad
\frac{E_{\rm FB}^{(\pm)}}{E_{\rm EM}^{(\pm)}}=\frac32.
\label{conclusion_branch_result}
\ee
Here $\Acal$ is the plate area and the superscripts $+$ and $-$ denote, respectively, equal and complementary ideal reflection, as defined in Section~\ref{section_fractonic_branch}. The simplicity of this result is itself informative: the Jordan enhancement of the unrestricted covariant determinant disappears because the vector modes responsible for it are absent from the physical fractonic branch.

The comparison also shows that reducing the bulk theory and reducing its boundary conditions are different operations. Natural boundary conditions derived for the full tensor field can overconstrain the three-mode branch because the Gauss constraint relates components that were independent before the reduction. For this reason the ideal branch reflectors were defined directly on the reduced physical space. A systematic construction of local surface actions compatible with that preferred-frame reduced phase space is an important extension; it would allow non-ideal and dispersive fractonic reflection to be studied without introducing arbitrary phenomenological coefficients.

Casimir-like attraction was previously found in a different fractonic setting, where Hilbert-space fragmentation generates an effective force between constrained excitations or between an excitation and a boundary \cite{Feng:2022fractoncasimir}. That result arises from the statistical structure of a many-body Hilbert space. The present calculation addresses a different problem: the vacuum energy of propagating higher-rank gauge-field modes modified by two boundaries. Its additional physical content comes from the comparison with the full covariant theory, which shows that the scalar-charge restriction does more than change the number of modes: it removes the unstable vector sector that would otherwise prevent a stable vacuum interpretation.

The same reduced action, varied as an autonomous theory, retains the transverse-vector amplitudes as a second and slower branch and gives the zero-temperature ratio $(3+\sqrt2)/2$ \cite{Muniz:2026fractoncasimir}. Imposing the branch on the covariant equations instead keeps the transverse projection of the Gauss constraint, removes exactly those two modes, and lowers the ratio by their contribution $\sqrt2/2$, so that what the covariant embedding does is visible as a definite number.

Other directions include the finite-temperature problem for the covariantly embedded branch and for the full higher-rank theory, which for the autonomous scalar-charge theory has very recently been analyzed in Ref.~\cite{Muniz:2026fractoncasimir}, together with curved boundaries, matter couplings and massive deformations. Covariant fracton theories on curved backgrounds have been constructed \cite{Afxonidis:2025curved}, while massive higher-rank models provide a guide to the crossover from power-law to exponentially suppressed vacuum interactions \cite{Bertolini:2025massive}. Dynamical boundary degrees of freedom may also connect the present slab problem with edge theories found in Maxwell and higher-rank gauge systems \cite{Bertolini:2020holographic,Bertolini:2023sqa,Bertolini:2023wie} and with the two-edge structure of Chern--Simons theory on a strip \cite{Bertolini:2026strip}.

The main physical message is therefore not a large numerical enhancement of a putative fracton Casimir force. The full covariant theory and the scalar-charge branch have different dynamical content. The Casimir problem makes that distinction measurable at the level of vacuum propagation: the unrestricted theory contains a coupled but Lorentzian-unstable vector sector, whereas the fractonic branch retains three stable massless modes whose ideal-reflector interaction is fixed relative to Maxwell by their physical multiplicity.

\appendix

\section{Mode sums, Euclidean determinants and the finite interaction}
\label{appendix_modesum}

For the stable Maxwell channels and for the three reduced channels of the fractonic branch, the finite determinant has the usual zero-point-energy interpretation. For a fixed tangential spatial momentum $k$, let $D_s(\omega,k;h)$ be the boundary determinant in a stable channel $s$. Its positive-frequency zeros are the allowed normal frequencies $\omega_n(k)$, and
\be
\frac{E(h)}{\Acal}
=\frac12\sum_s\int\frac{\dd^2k}{(2\pi)^2}\sum_n\omega_n(k).
\label{app_modesum}
\ee
The argument principle converts the sum over zeros into a contour integral,
\be
\sum_n\omega_n
=\frac{1}{2\pi i}\oint_{\mathcal C}\dd\omega\,
\omega\,\partial_\omega\log D_s(\omega,k;h),
\label{app_argument}
\ee
where $\mathcal C$ encloses the positive-frequency zeros \cite{Lambrecht:2006few,Rahi:2009hm}. Rotating the contour to imaginary frequency, integrating by parts and combining the imaginary frequency with the two tangential momenta gives
\be
\frac{E(h)}{\Acal}
=\frac12\sum_s\int\frac{\dd^3p}{(2\pi)^3}\log D_s(q,h)
+\text{local terms}.
\label{app_euclidean}
\ee
The local terms are independent of $h$ or proportional to the slab volume. They are removed by the subtraction in Eq.~\eqref{subtraction}. Equivalently, one divides the determinant by the factors that remain when the two plates are infinitely separated. For one independent stable channel this gives
\be
\Delta_{\rm H}(X)=1-r_0r_h e^{-2X}.
\label{app_scalar_factor}
\ee
This derivation applies directly to both photon polarizations and to the three physical channels of the fractonic branch.

The Jordan doublets of the full covariant theory require a different interpretation. Section~\ref{section_parent_vector_stability} shows that their Lorentzian time evolution has repeated roots, generalized solutions and an indefinite Hamiltonian. There is therefore no stable positive-frequency spectrum for which Eq.~\eqref{app_modesum} defines a ground-state energy. Their Euclidean quantities are instead defined directly by the Gaussian functional integral of the positive Euclidean quadratic action. At fixed $p_a$, the four boundary conditions form the secular matrix $\mathcal S_s(q,h)$ of Eq.~\eqref{secular_system}; its determinant gives the Gaussian boundary factor. Dividing by the large-separation factors yields
\be
\Delta_{1,s}(q,h)
=\frac{\det\mathcal S_s(q,h)}{\det\mathcal S_s(q,\infty)}\Bigg|_{\rm finite},
\label{app_parent_euclidean_factor}
\ee
Its contribution to the normalized Euclidean unrestricted functional is
\be
\frac12\int\frac{\dd^3p}{(2\pi)^3}\,\log\Delta_{1,s}.
\label{app_jordan_functional}
\ee
The algebraic subtraction is the same one used in the stable mode-sum problem, but no Lorentzian zero-point-energy interpretation is assigned to the full-theory Jordan factor.


\section{Gauge-sector boundary problem and determinant cancellation}
\label{appendix_gauge_sector}

\subsection{Variational rederivation without ghosts}
\label{appendix_variational_gauge}

The boundary domain of the gauge sector and the cancellation of the non-physical determinants were obtained above from the local BRS-exact functional \eqref{boundary_gauge_fermion}. This subsection gives a complementary derivation of the natural boundary conditions using the classical gauge-fixed action alone. It takes the same essential boundary gauge slice \eqref{longitudinal_essential_bc} as input. The calculation is valid for the full gauge family at the level of the boundary variation. The explicit determinant cancellation is treated separately in Appendix~\ref{appendix_gauge_determinant} in the convenient gauge $\chi=0$, where the transverse trace decouples from the longitudinal block.

\subsubsection{Essential and natural boundary data}

Take as essential boundary data on the longitudinal block the conditions of Eq.~\eqref{longitudinal_essential_bc}. This local boundary gauge slice selects a representative of the gauge orbit at the surfaces. The physical response of the plates is carried independently by the surface action of Section~\ref{section_boundary_doublets}. The remaining free variations determine the natural conditions.

The bulk action \eqref{tensor_action} contains at most first normal derivatives, so it contributes only to the coefficients of $\delta A_{\mu\nu}$; the coefficients of $\delta A'_{\mu\nu}$ come entirely from the gauge-fixing term. Varying $S_{\rm F}+S_{\rm gf}$ with $\delta A_{pp}=\delta A_{p3}=0$ on the plates gives the boundary coefficients
\be
\begin{aligned}
\text{coeff}\left(\delta A'_{pp}\right)&=\frac{\chi g_{\rm F}}{\xi}\ G_\chi,
&\qquad
\text{coeff}\left(\delta A'_{33}\right)&=\frac{\left(1+\chi\right)g_{\rm F}}{\xi}\ G_\chi,\\
\text{coeff}\left(\delta A'_{p3}\right)&=0,
&\qquad
\text{coeff}\left(\delta A_{33}\right)&=-\frac{\left(1+\chi\right)g_{\rm F}}{\xi}\ \partial_3G_\chi.
\end{aligned}
\label{app_var_coefficients}
\ee
These algebraic identities hold before the boundary conditions are applied. At $\chi=0$ the coefficient of $\delta A'_{pp}$ vanishes, but the coefficient of $\delta A'_{33}$ remains nonzero. Therefore, for every $\chi\neq-1$, the free variations produce exactly the natural conditions of Eq.~\eqref{longitudinal_natural_bc}, now obtained without any use of the BRS construction. Together with the essential data these are four conditions per plate, the number required by the fourth-order longitudinal problem. At $\chi=-1$, the coefficient multiplying the gauge variation vanishes and $G_{-1}$ becomes gauge invariant; the gauge family therefore covers $\chi\neq-1$. The value $\chi=0$ is regular and is precisely the convenient gauge used in the explicit calculation above.

\subsubsection{Scope of the variational check}

For $\chi\neq0$, the off-shell gauge-fixing term $G_\chi^2$ mixes the longitudinal variables with the gauge-invariant transverse trace. A determinant obtained by setting the transverse trace to zero is therefore not the determinant of a closed block and is not used as evidence for spectral factorization at arbitrary $\chi$. The gauge independence of the complete normalized functional follows instead from the BRS identities in Eq.~\eqref{gauge_parameter_independence_full}. The explicit spectral cancellation below is evaluated at $\chi=0$, where the trace and longitudinal sectors separate exactly.

\subsection{BRS boundary domain and explicit Faddeev--Popov cancellation}
\label{appendix_gauge_determinant}

This subsection supplies the details used in Sections~\ref{section_brs_boundary} and \ref{section_gauge_determinant}. We first verify that the boundary domain follows from a local BRS-exact completion and is closed off shell. We then compute the longitudinal and Faddeev--Popov determinants in the convenient gauge $\chi=0$.

\subsubsection{Boundary gauge-fixing fermion and off-shell closure}

The two boundary gauge functions in Eq.~\eqref{boundary_gauge_functions} transform as
\be
s\mathcal F_{1,\alpha}
=(\partial_\parallel^2)^2c\big|_{\Sigma_\alpha},
\qquad
s\mathcal F_{2,\alpha}
=\partial_\parallel^2\partial_3c\big|_{\Sigma_\alpha}.
\label{app_boundary_gauge_variations}
\ee
Equation~\eqref{boundary_brs_action} follows directly from Eq.~\eqref{boundary_brs_doublets}. For $q>0$, integration over the boundary multipliers $b_\alpha,d_\alpha$ imposes $A_{pp}=A_{p3}=0$, while integration over $\bar\eta_\alpha,\bar\vartheta_\alpha$ imposes $c=c'=0$. The boundary fields carry no normal kinetic term. Their Gaussian or multiplier factors depend on $q$ but not on $h$ and therefore cancel from the interaction normalization.

The Green identity for the fourth-order scalar operator
\be
L_q^2=(\partial_3^2-q^2)^2
\label{app_ghost_operator}
\ee
is
\be
\begin{aligned}
\int_0^h\dd x^3\,
\left[\bar c L_q^2c-(L_q^2\bar c)c\right]
=\Big[
\bar c(L_qc)' -\bar c'(L_qc)
+(L_q\bar c)c'-(L_q\bar c)'c
\Big]_{0}^{h}.
\end{aligned}
\label{app_ghost_green_identity}
\ee
It vanishes for the clamped domains $c=c'=0$ and $\bar c=\bar c'=0$. The antighost domain is therefore the adjoint domain of the ghost operator. Since $s\bar c=B$, its BRS image is precisely $B=B'=0$. The complete set is closed off shell:
\be
\begin{aligned}
&s(A_{pp})=-q^2c=0,
&&s(A_{p3})=iqc'=0,\\
&s c=0,
&&s\bar c=B=0,\\
&s(\bar c')=B'=0,
&&sB=0,
\end{aligned}
\qquad\text{at }x^3=0,h.
\label{app_offshell_closure}
\ee
No field equation has been used in Eq.~\eqref{app_offshell_closure}.

The first-order gauge-fixing action contains $BG_\chi$. Varying it with respect to the full tensor field and integrating normal derivatives by parts produces boundary terms proportional to $B$, $B'$ and to variations of the tensor field and its first normal derivative. They vanish on Eq.~\eqref{B_clamped_domain}. On the clamped ghost domain, the BRS variation of every component entering the physical surface action vanishes at the plates; the physical surface action is therefore BRS invariant on this boundary domain and supplies the independent physical boundary response. Consequently the BRS completion neither changes nor supplements those physical conditions.

Varying with respect to $B$ gives Eq.~\eqref{B_equation}. Only after this algebraic equation is used do $B=B'=0$ become $G_\chi=G_\chi'=0$. In the off-shell domain the independent entries are the tensor, ghost, antighost and multiplier conditions; the gauge-function conditions arise after the algebraic multiplier equation is used.

For completeness, the parameter identities follow directly from the gauge-fixing fermion:
\be
\frac{\partial}{\partial\xi}s\Psi_{\xi,\chi}^{\rm bulk}
=s\left(-\frac{g_{\rm F}}2\int_\M\dd^4x\,\bar cB\right),
\qquad
\frac{\partial}{\partial\chi}s\Psi_{\xi,\chi}^{\rm bulk}
=s\left(g_{\rm F}\int_\M\dd^4x\,\bar c\Box A\right).
\label{app_parameter_exactness}
\ee
Because the integration domain is BRS invariant, the expectation value of either BRS-exact insertion vanishes. This proves Eq.~\eqref{gauge_parameter_independence_full} for every plate separation before the Casimir subtraction is taken.

\subsubsection{Longitudinal operator in the convenient gauge}

Let $\Phi$ be the longitudinal triple of Eq.~\eqref{longitudinal_vector}. After the equation for the off-diagonal component $A_{p3}$ is multiplied by two, an $h$-independent normalization convenient for the quadratic form, the ungauge-fixed longitudinal operator evaluated on $e^{\kappa x^3}$ is
\be
\mathcal H_0(\kappa)=
\begin{pmatrix}
-2\kappa^2 & 2iq\kappa & 0\\
2iq\kappa & 2(q^2-\kappa^2) & 2iq\kappa\\
0 & 2iq\kappa & 2q^2
\end{pmatrix}.
\label{app_H0}
\ee
The gauge direction and the symbol of $G=G_0$ are
\be
 d(\kappa)=
\begin{pmatrix}-q^2\\iq\kappa\\\kappa^2\end{pmatrix},
\qquad
 g(\kappa)=
\begin{pmatrix}-q^2\\2iq\kappa\\\kappa^2\end{pmatrix},
\label{app_d_g_vectors}
\ee
so that $G=g^{T}\Phi$ for the exponential ansatz. Direct multiplication gives
\be
\mathcal H_0d=0,
\qquad
g^{T}d=(\kappa^2-q^2)^2.
\label{app_noether_symbol}
\ee
The first identity is the Noether identity in the longitudinal block; the second is the Faddeev--Popov operator evaluated on a gauge transformation.

After $B$ is eliminated, the Euclidean gauge-fixed operator is the rank-one completion
\be
N_\xi(\kappa)
=\mathcal H_0(\kappa)+\frac1\xi
g(\kappa)g^{T}(\kappa).
\label{app_general_N}
\ee
Its determinant is
\be
\det N_\xi(\kappa)
=\frac4\xi(\kappa^2-q^2)^4.
\label{app_general_detN}
\ee
Thus $\kappa=\pm q$ are roots of multiplicity four and the normal solution space has dimension eight.

In coordinate space,
\be
G=-q^2A_{pp}+2iqA_{p3}'+A_{33}''.
\label{app_G_coordinate}
\ee
The Green boundary form of the gauge-invariant operator vanishes on $A_{pp}=A_{p3}=0$. In the first-order formulation the remaining gauge-fixing boundary form is killed by $B=B'=0$. Eliminating $B$ then gives
\be
\left.\delta S_{\rm gf}\right|_{\partial\M}
=\frac{g_{\rm F}}{\xi}
\sum_{\alpha=0,h}\sigma_\alpha
\left[G^*\delta A_{33}'-(G^*)'\delta A_{33}\right]+\text{c.c.},
\label{app_boundary_form}
\ee
which vanishes because $G=G'=0$ is the on-shell image of the multiplier domain. No additional condition is introduced at this stage.

\subsubsection{Exact sequence and secular determinants}

Write
\be
L_q\equiv(\partial_3^2-q^2)^2,
\qquad
\mathcal K_q\equiv\ker L_q.
\label{app_Lq}
\ee
For $q>0$, $\mathcal K_q$ is four-dimensional and is spanned by the profiles in Eq.~\eqref{FP_normal_basis}. The differential gauge generator
\be
D\Lambda=
\begin{pmatrix}-q^2\Lambda\\iq\Lambda'\\\Lambda''\end{pmatrix}
\label{app_D_operator}
\ee
embeds $\mathcal K_q$ into the solution space $\mathcal S_N$ of $N_\xi\Phi=0$. Indeed, $\mathcal H_0D=0$ and $G[D\Lambda]=L_q\Lambda$. Conversely, applying the adjoint gauge generator to the gauge-fixed field equation removes the gauge-invariant part by the Noether identity and gives
\be
L_qG=0.
\label{app_LqG}
\ee
If a solution satisfies $G=0$, it obeys the ungauge-fixed longitudinal equation. The kernel of $\mathcal H_0(\kappa)$ is generated by $d(\kappa)$, including its generalized profiles at $\kappa=\pm q$, so this solution is pure gauge. Equation~\eqref{app_general_detN} gives $\dim\mathcal S_N=8$; the four-dimensional quotient is therefore mapped onto $\mathcal K_q$ by $G$. The result is the exact sequence
\be
0\longrightarrow\mathcal K_q
\overset{D}{\longrightarrow}\mathcal S_N
\overset{G}{\longrightarrow}\mathcal K_q
\longrightarrow0.
\label{app_exact_sequence}
\ee
This is the precise form of the decomposition $8=4+4$ used in the main text.

Choose four pure-gauge columns $Df_j$, where $f_j$ is a basis of $\mathcal K_q$, and four complementary columns $W_j$ satisfying $G[W_j]=f_j$. Order the boundary rows as
\be
\left(A_{pp},A_{p3}\right)_{0,h}
\quad\text{followed by}\quad
\left(G,G'\right)_{0,h}.
\label{app_row_order}
\ee
On the pure-gauge columns the lower rows vanish, whereas the upper rows are the clamped boundary matrix of $f_j$, multiplied by powers of $q$ independent of $h$. On the complementary columns the lower block is the same clamped matrix because $G[W_j]=f_j$. Hence
\be
\mathcal S_N=
\begin{pmatrix}
\mathcal Q\mathcal S_{\rm FP}& *\\
0&\mathcal S_{\rm FP}
\end{pmatrix},
\qquad
D_N=C(q,\xi)D_{\rm FP}^2,
\label{app_block_triangular}
\ee
where $\det\mathcal Q$ and $C$ are independent of $h$.

Using the dimensionless coordinate $y=qx^3$, the clamped scalar boundary matrix is
\be
\mathcal S_{\rm FP}(X)=
\begin{pmatrix}
1&0&1&0\\
1&1&-1&1\\
e^X&Xe^X&e^{-X}&Xe^{-X}\\
e^X&(1+X)e^X&-e^{-X}&(1-X)e^{-X}
\end{pmatrix},
\label{app_FP_matrix}
\ee
with
\be
D_{\rm FP}(X)=\det\mathcal S_{\rm FP}
=4\left(\sinh^2X-X^2\right).
\label{app_FP_det}
\ee
Dividing by its large-$X$ behavior gives Eq.~\eqref{FP_interaction_factor}. The block-triangular identity then gives Eq.~\eqref{normalized_longitudinal_factorization} for arbitrary $\xi>0$ in the gauge $\chi=0$. Gauge independence with respect to both $\xi$ and $\chi$ follows separately from Eq.~\eqref{gauge_parameter_independence_full}. Appendix~\ref{appendix_variational_gauge} checks the associated natural boundary conditions at arbitrary $\chi\neq-1$; it is not used as an arbitrary-$\chi$ determinant factorization.

As a direct check, set $q=\xi=1$. Equation~\eqref{app_general_N} becomes
\be
N_1(\kappa)=
\begin{pmatrix}
1-2\kappa^2 & 0 & -\kappa^2\\
0 & 2-6\kappa^2 & 2i\kappa(\kappa^2+1)\\
-\kappa^2 & 2i\kappa(\kappa^2+1) & \kappa^4+2
\end{pmatrix}.
\label{app_representative_N}
\ee
Constructing its eight generalized profiles and applying the four conditions at each plate gives
\be
D_N(X)=-589824\left(\sinh^2X-X^2\right)^2
=-36864\,D_{\rm FP}^2(X).
\label{app_explicit_factorization}
\ee
The numerical coefficient refers to the particular normalization chosen for the generalized solution basis in this calculation. It is independent of $h$ and therefore disappears from the normalized interaction determinant; the normalization-independent content is the identity $\Delta_N=\Delta_{\rm FP}^2$. 

At $q=0$, the polarization vector $\widehat p^\mu$ and the boundary gauge functions do not resolve the scalar gauge orbit. The zero-momentum gauge sector must therefore be separated before the helicity decomposition and divided by its residual gauge volume. It is a single point in the $\dd^3p$ integral and does not contribute to the normalized interaction density.


\section{Algebra of the full-theory normal modes}
\label{appendix_modes}

\subsection{Fourier convention and bulk symbol}

The tangential Fourier transform is
\be
A_{\mu\nu}(x^a,x^3)
=\int\frac{\dd^3p}{(2\pi)^3}\,
e^{ip_ax^a}\widetilde A_{\mu\nu}(p,x^3).
\label{app_fourier}
\ee
Choose the tangential momentum along one tangential axis and insert Eq.~\eqref{normal_ansatz}. In a convenient Euclidean basis, derivatives act as
\be
\partial_\mu\longrightarrow k_\mu=(0,0,iq,\kappa),
\qquad
k^2=\kappa^2-q^2.
\label{app_k}
\ee
The field equations become
\be
-2k^2A_{\mu\nu}+k_\mu B_\nu+k_\nu B_\mu=0,
\qquad
B_\nu=k^\rho A_{\nu\rho}.
\label{app_symbol}
\ee
The contractions orthogonal to both $k_\mu$ and the normal give the three Helmholtz equations in Eq.~\eqref{tensor_helmholtz}. The $s=\pm1$ contractions give Eq.~\eqref{doublet_matrix}.

\subsection{Generalized solutions of one doublet}

At $\kappa=\pm q$, the kernel of $\mathcal M_1$ is one-dimensional. An ordered basis of four normal solutions is
\be
\varphi_1=
\begin{pmatrix}i\\1\end{pmatrix}e^{qx^3},
\qquad
\varphi_2=
\begin{pmatrix}ix^3\\x^3+3/q\end{pmatrix}e^{qx^3},
\label{app_basis_plus}
\ee
and
\be
\varphi_3=
\begin{pmatrix}-i\\1\end{pmatrix}e^{-qx^3},
\qquad
\varphi_4=
\begin{pmatrix}-ix^3\\x^3-3/q\end{pmatrix}e^{-qx^3}.
\label{app_basis_minus}
\ee
Direct substitution verifies Eq.~\eqref{doublet_equations}.

\section{Local surface action and explicit doublet determinants}
\label{appendix_surface}

\subsection{A representative local tensor action}

A representative action in the minimal local surface class is
\be
\begin{aligned}
S_{\partial}=\sum_{\alpha=0,h}\int_{\Sigma_\alpha}\dd^3x\,\bigg[&
\frac{\lambda_{T,\alpha}}2A_{ab}A^{ab}
+\frac{\lambda_{{\rm tr},\alpha}}2(A^a{}_{a})^2
+\lambda_{N,\alpha}A_{a3}A^{a3}\\
&+\eta_{1,\alpha}A^{ab}\partial_aA_{b3}
+\eta_{2,\alpha}A^a{}_{a}\partial^bA_{b3}\\
&+\frac{\gamma_{T,\alpha}}2\epsilon^{abc}A_a{}^d\partial_bA_{cd}
+\frac{\gamma_{N,\alpha}}2\epsilon^{abc}A_{a3}\partial_bA_{c3}
\bigg].
\end{aligned}
\label{app_surface_action}
\ee
Here $\epsilon^{abc}$ is the plate-coordinate form of the induced orientation tensor in Eq.~\eqref{boundary_geometry}. The $\lambda$ coefficients have mass dimension one, while $\eta$ and $\gamma$ are dimensionless. Projection of Eq.~\eqref{app_surface_action} onto a fixed $s=\pm1$ doublet gives the matrix in Eq.~\eqref{response_matrix}. Different tensor actions can lead to the same projected matrix; the normalized unrestricted interaction depends on the projected response of the Jordan doublet.

\subsection{Dirichlet determinant}

Using the basis in Eqs.~\eqref{app_basis_plus} and \eqref{app_basis_minus}, the Dirichlet conditions form the matrix
\be
\mathcal S_{\rm D}(h)=
\begin{pmatrix}
i & 0 & -i & 0\\
1 & 3/q & 1 & -3/q\\
iE & ihE & -iE^{-1} & -ihE^{-1}\\
E & (h+3/q)E & E^{-1} & (h-3/q)E^{-1}
\end{pmatrix},
\qquad
E=e^{qh}.
\label{app_dirichlet_matrix}
\ee
Its determinant is
\be
\det\mathcal S_{\rm D}(h)
=-\frac4{q^2}\left[9\sinh^2X-X^2\right].
\label{app_dirichlet_det}
\ee
Using
\be
\sinh^2X=\frac14e^{2X}\left(1-2e^{-2X}+e^{-4X}\right),
\label{app_sinh}
\ee
and removing the factor $-9e^{2X}/q^2$ that survives in the isolated-plate limit gives Eq.~\eqref{deltaD}. The same determinant can be obtained with the matrix form of the Gelfand--Yaglom method \cite{Forman:1987,KirstenMcKane:2004}.

\subsection{Free, mixed and general scale-free response}

It is useful to display the matrices leading to the other two representative determinants. We set
\be
E\equiv e^X
\label{app_E_definition}
\ee
and remove row and column factors that depend on $q$ but not on $h$, because they cancel in the normalized interaction determinant.

For free boundary conditions on both plates, the four equations in Eq.~\eqref{free_bc} give
\be
\mathcal S_{\rm free}(X)=
\begin{pmatrix}
 i &-i&i&i\\
 1&4&-1&4\\
 iE&i(X-1)E&iE^{-1}&i(X+1)E^{-1}\\
 E&(X+4)E&-E^{-1}&(4-X)E^{-1}
\end{pmatrix}.
\label{app_free_matrix}
\ee
Its determinant is
\be
\det\mathcal S_{\rm free}
=4X^2-25E^2+50-25E^{-2}.
\label{app_free_det}
\ee
Dividing by the leading isolated-plate factor $-25E^2$ yields Eq.~\eqref{deltafree}.

For Dirichlet conditions at $\Sigma_0$ and free conditions at $\Sigma_h$, the corresponding matrix is
\be
\mathcal S_{\rm mix}(X)=
\begin{pmatrix}
 i&0&-i&0\\
 1&3&1&-3\\
 iE&i(X-1)E&iE^{-1}&i(X+1)E^{-1}\\
 E&(X+4)E&-E^{-1}&(4-X)E^{-1}
\end{pmatrix}.
\label{app_mixed_matrix}
\ee
Its determinant is
\be
\det\mathcal S_{\rm mix}
=-15E^2-(4X^2+34)-15E^{-2}.
\label{app_mixed_det}
\ee
After division by $-15E^2$, this gives Eq.~\eqref{deltamixed}.

The same calculation can be carried out for a scale-free diagonal response. For identical plates, define
\be
\widehat a=\frac{a_\alpha}{q},
\qquad
\widehat c=\frac{c_\alpha}{q}.
\label{app_ahat_chat}
\ee
The secular determinant has three powers of $E$,
\be
D(X)=A_2E^2+A_0(X)+A_{-2}E^{-2}.
\label{app_three_powers}
\ee
Dividing by the leading term gives
\be
\Delta_1(X)
=1+\frac{A_0(X)}{A_2}e^{-2X}
+\frac{A_{-2}}{A_2}e^{-4X}.
\label{app_general_family}
\ee
The coefficients are
\be
A_2=-\left(3\widehat a\widehat c+4\widehat a+4\widehat c+5\right)^2,
\qquad
A_{-2}=-\left(3\widehat a\widehat c-4\widehat a-4\widehat c+5\right)^2,
\label{app_Apm}
\ee
and
\be
\begin{aligned}
A_0(X)=2\big[&2(\widehat a^2\widehat c^2-2\widehat a\widehat c+1)X^2
+16(\widehat a-\widehat c)(\widehat a\widehat c-1)X\\
&+9\widehat a^2\widehat c^2+16\widehat a^2+16\widehat c^2
-66\widehat a\widehat c+25\big].
\end{aligned}
\label{app_A0}
\ee
Comparison with Eq.~\eqref{general_delta} identifies
\be
P(X)=-\frac{A_0(X)}{A_2},
\qquad
Q=\frac{A_{-2}}{A_2}.
\label{app_PQ_identification}
\ee

\section{Orientation-dependent determinants and boundary stability}
\label{appendix_stability}

\subsection{Diagonal coupling}

For the coefficients in Eq.~\eqref{gamma_surface_coefficients}, the four boundary equations are those displayed in Eq.~\eqref{gamma_boundary_conditions}. Evaluating them on the basis in Eqs.~\eqref{app_basis_plus} and \eqref{app_basis_minus} gives a determinant of the form
\be
D_\gamma(X)=C_\gamma(q)\,(1-\gamma^2)(25-9\gamma^2)e^{2X}
\left[1-P_\gamma(X)e^{-2X}+e^{-4X}\right],
\label{app_gamma_det}
\ee
where $C_\gamma(q)$ is independent of $h$ and $P_\gamma(X)$ is the coefficient of Eq.~\eqref{Pgamma}. The prefactor in Eq.~\eqref{app_gamma_det} belongs to the isolated-plate normalization. Removing it gives Eqs.~\eqref{delta_gamma_explicit}--\eqref{delta_gamma}.

The half-space boundary-stability test starts from the eigenvalue problem \eqref{surface_eigenvalue_problem}. For $\gamma>0$, the decaying profile \eqref{surface_mode} belongs to the $s=-1$ doublet at the lower plate. Direct substitution gives
\be
\mathcal M_1(\partial_3)
\begin{pmatrix}-i\\ \gamma\end{pmatrix}e^{-\gamma qx^3}
=q^2(1-\gamma^2)
\begin{pmatrix}-i\\ \gamma\end{pmatrix}e^{-\gamma qx^3},
\label{app_gamma_eigenvalue_check}
\ee
and the same profile satisfies the two boundary equations at the isolated plate. This proves Eq.~\eqref{surface_eigenvalue}.

\subsection{Off-diagonal coupling}

For the coefficients in Eq.~\eqref{beta_surface_coefficients}, the four equations are given in Eq.~\eqref{beta_boundary_conditions}. Their secular determinant has coefficients
\be
\begin{aligned}
A_2&=-(\beta-1)^2(3\beta+5)^2,\\
A_0(X)&=2(\beta-1)^2\left[2(\beta-1)^2X^2+(3\beta+5)^2\right],\\
A_{-2}&=A_2.
\end{aligned}
\label{app_beta_coefficients}
\ee
Using Eq.~\eqref{app_general_family} gives
\be
\Delta_\beta(X)
=1-\left[2+4X^2\frac{(\beta-1)^2}{(3\beta+5)^2}\right]e^{-2X}+e^{-4X},
\label{app_beta_delta_explicit}
\ee
which is equivalent to Eqs.~\eqref{delta_beta} and \eqref{Nbeta}.

To analyze an isolated plate, write a possible negative eigenvalue as
\be
\lambda=-\mu q^2,
\qquad
\mu>0,
\label{app_negative_eigenvalue}
\ee
and define the two positive decay constants
\be
r=\sqrt{1+\mu},
\qquad
\sigma=\sqrt{1+\frac{\mu}{2}}.
\label{app_r_sigma}
\ee
Evaluating the boundary conditions on the two independent decaying solutions gives the half-space boundary determinant
\be
D_\beta^{\rm half}(\mu)
=\frac{(r^2-\beta)^2}{2r}
-\frac{(1-\beta)^2\sigma}{2}.
\label{app_halfspace_det}
\ee
At $\mu=0$ the two bulk solutions coalesce, producing a kinematic factor $r-\sigma$. The normalized limit is
\be
\lim_{\mu\to0^+}
\frac{D_\beta^{\rm half}(\mu)}{r-\sigma}
=\frac12(1-\beta)(3\beta+5).
\label{app_halfspace_beta}
\ee
For $-5/3<\beta<1$, the determinant remains positive for every $\mu>0$. One way to see this is to write
\be
2rD_\beta^{\rm half}
=(r^2-\beta)^2-r\sigma(1-\beta)^2
\label{app_halfspace_factor}
\ee
and use the lower bound
\be
r^2+\frac53-\frac83\sqrt{r\sigma}>0,
\qquad r>1.
\label{app_halfspace_bound}
\ee
The bound follows from
\be
(3r^2+5)^4-2048r^2(r^2+1)
=(r^2-1)^2\left(81r^4+702r^2+625\right)>0.
\label{app_beta_identity}
\ee
Outside this interval the normalized determinant has the opposite sign near $\mu=0$ and becomes positive for large $\mu$, so a negative surface eigenvalue is present. This establishes Eq.~\eqref{beta_oneplate}.

For the two-plate problem, Eq.~\eqref{delta_beta} can be rewritten as
\be
\Delta_\beta(X)
=4e^{-2X}\left[\sinh^2X-\frac{X^2}{N_\beta^2}\right].
\label{app_beta_sinh}
\ee
Within the half-space interval, $N_\beta=(3\beta+5)/(1-\beta)>0$. Since $\sinh X>X$ for $X>0$, the determinant is positive for every $X>0$ when $N_\beta\geq1$. The small-$X$ expansion shows that this condition is also necessary. Thus
\be
N_\beta\geq1
\quad\Longleftrightarrow\quad
\beta\geq-1,
\label{app_beta_positivity}
\ee
and intersection with Eq.~\eqref{beta_oneplate} gives the slab domain \eqref{beta_slab}.

\end{document}